\documentclass[sinlgecolumn,preprintnumbers,amsmath,amssymb,superscriptaddress,longbibliography,nofootinbib, notitlepage]{revtex4-1}

\usepackage{graphicx}
\usepackage{bm}
\usepackage{dsfont}
\usepackage[usenames,dvipsnames]{xcolor}
\usepackage{pstricks}
\usepackage[tight]{subfigure}
\usepackage{verbatim}
\usepackage{units}
\usepackage{multirow}
\usepackage{enumitem}
\usepackage{mathrsfs}
\usepackage{leftidx}
\usepackage{xspace}
\usepackage{braket}
\usepackage{bbm}
\usepackage{cancel}
\usepackage{amsfonts,amssymb,amsmath}
\usepackage[normalem]{ulem}

\usepackage{xr}

\usepackage[a4paper]{hyperref}
\hypersetup{colorlinks=true,linktoc=all,linkcolor=blue,breaklinks=true,citecolor=blue,urlcolor=blue}
\usepackage[english]{babel}
\usepackage{cleveref}

\newcommand{\kB}{k_\text{B}}

\begin{document}

\title{Combining kinetic and thermodynamic uncertainty relations in quantum transport}

\author{Didrik Palmqvist}
	\affiliation{Department of Microtechnology and Nanoscience (MC2), Chalmers University of Technology, S-412 96 G\"oteborg, Sweden}

\author{Ludovico Tesser}
	\affiliation{Department of Microtechnology and Nanoscience (MC2), Chalmers University of Technology, S-412 96 G\"oteborg, Sweden}

	\author{Janine Splettstoesser}
	\affiliation{Department of Microtechnology and Nanoscience (MC2), Chalmers University of Technology, S-412 96 G\"oteborg, Sweden}
	
\date{\today}

\begin{abstract}
 We study the fluctuations of generic currents in multi-terminal, multi-channel quantum transport settings. In the quantum regime, these fluctuations and the resulting precision differ strongly depending on whether the device is of fermionic or bosonic nature. Using scattering theory, we show that the precision is bounded by constraints set by the entropy production and by the activity in the spirit of thermodynamic or kinetic uncertainty relations, valid for fermionic and bosonic quantum systems and even in the absence of time-reversal symmetry. Furthermore, we derive a combined thermodynamic kinetic uncertainty relation, which is tight over a wide range of parameters and can hence predict the reachable precision of a device. 
 Since these constraints can be expressed in terms of observables accessible in transport measurements, such as currents and bandwidth, we foresee that the tight thermodynamic kinetic uncertainty-like bounds are also useful as an inference tool: they can be exploited to estimate entropy production from transport observables, such as the charge current and its noise, which are more easily accessible in experiment.  
\end{abstract}

\maketitle

\section{Introduction}

In small-scale devices, fluctuations around an average signal can be of significant magnitude~\cite{Blanter2000Sep}. Consequently, achieving high precision---or equivalently achieving a high signal-to-noise ratio---is crucial in order to access the actual signal in an experiment or to produce a reliable outcome of a process. In quantum transport, such a desired outcome could be a precise charge current in response to a voltage or reliable electrical or chemical power produced by exploiting some available resource, such as heat~\cite{Benenti2017Jun,Pekola2021Oct}.

In order to optimize processes with respect to precision, it is important to know what the maximum achievable precision is and what constrains it. Close to equilibrium~\cite{Callen1951Jul,Kubo1957Jun} or under specific nonequilibrium conditions~\cite{Rogovin1974Jul,Levitov2004Sep,Altaner2016Oct,Dechant2020Mar,Shiraishi2022Jul}, general relations between a current and its noise can be found in terms of fluctuation-dissipation theorems. Also, if the explicit model of the system is known, the average current and its noise can ``simply" be calculated from appropriate transport theory methods. 
To make generally valid predictions about the relationship between currents and noise under nonequilibrium conditions and without knowledge about the microscopic model of the device, several \textit{bounds} on precision have recently been developed~\cite{Landi2024Apr}. 
These inequalities, originally developed for classical Markovian processes, constrain precision in terms of entropy production, known as the thermodynamic uncertainty relation (TUR)~\cite{Barato2015Apr,Gingrich2016Mar}, or in terms of activity, known as the kinetic uncertainty relation (KUR)~\cite{DiTerlizzi2018Dec,Liu2025Feb}. While the TUR is most predictive close to equilibrium, the bound set by the KUR is most tight far from equilibrium, and recently a unified thermodynamic kinetic uncertainty relation (TKUR) was derived, establishing a crossover between the TUR and KUR while providing an overall tighter bound on precision~\cite{Vo2022Sep}.

\begin{figure}[b!]
    \centering
   \includegraphics[width=3.3in]{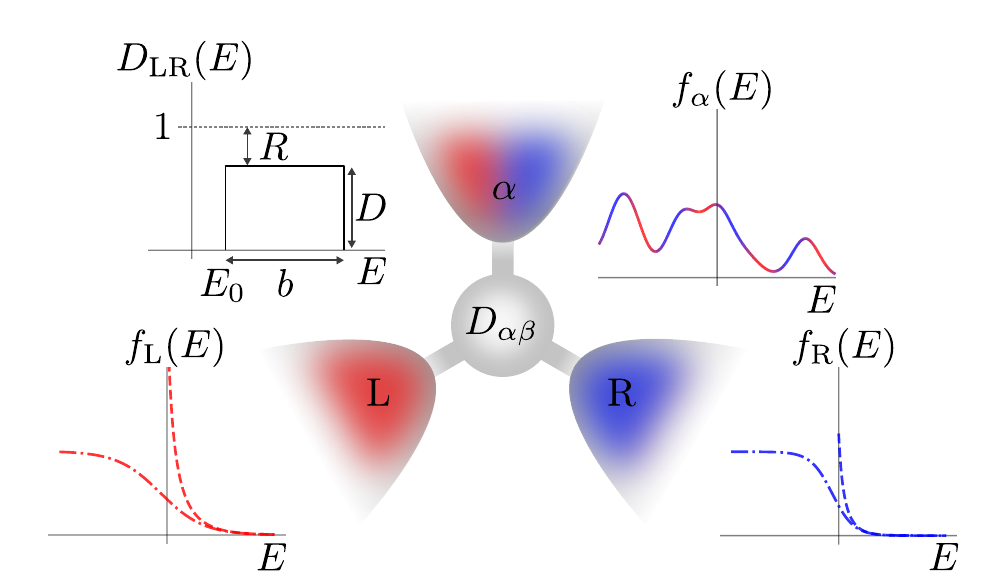}
    \caption{Sketch of a multi-terminal quantum transport setup. The central scattering region is characterized by multi-channel scattering submatrices $t_{\alpha\beta}(E)$ through which the transmission function $D_{\alpha \beta}(E)=\text{Tr}\{t_{\alpha\beta}^\dagger(E) t_{\alpha\beta}(E)\}$ is defined. Contacts are occupied following Fermi or Bose distributions, or even nonthermal distributions, which can not be characterized by a temperature or a chemical potential. The lower two reservoirs depict a thermal two-terminal system connected by a boxcar transmission $D_{\mathrm{LR}}(E)$ for which we demonstrate our bounds. }
    \label{fig:setup}
\end{figure}
Lately, considerable efforts have been made to include the impact of quantum effects in these bounds~\cite{Hasegawa2024Dec}, mostly using quantum master equations and, in particular, master equations unraveled into quantum jump operators~\cite{VanVu2022Apr,Hasegawa2023May,Prech2023Jun,Manzano2023Oct,Kwon2024Dec,Prech2025Jan,Yunoki2025Feb}. 
A complementary approach to treating quantum effects in transport is by using scattering theory \cite{Blanter2000Sep,Moskalets2011Sep}. This has the advantage of fully treating quantum effects at the cost of treating many-body interactions at most at the mean-field level~\cite{Jauho1994Aug,Souza2008Oct,Haupt2010Oct}, while still allowing inelastic effects to be modelled through B\"uttiker probes~\cite{Texier2000Sep}.
Furthermore, while the definition of activity is not clearly established when going beyond classical stochastic processes, a treatment by scattering theory allows one to establish bounds in terms of measurable quantities (like noise, transport bandwidth, etc.), only.
Using scattering theory for transport, fluctuation-dissipation bounds~\cite{Tesser2024May} and TURs~\cite{Brandner2018Mar,Potanina2021Apr,Brandner2025Feb} have been found for fermionic systems, and KURs have been established both for the bosonic and fermionic case~\cite{Palmqvist2024Oct}. However, a transport TKUR does to our knowledge not exist so far, and also TURs for bosons in systems with broken time-reversal symmetry have not been studied.

In this paper, we fill these gaps and provide precision bounds for transport in the form of TURs for arbitrary types of contacts and in the form of KUR- and TKUR-like bounds. To derive these bounds, we use scattering theory for multi-terminal, multi-channel bosonic or fermionic systems. We allow for broken time-reversal symmetry and even include the case of ``nonthermal" distributions in the contacts~\cite{Sanchez2019Nov,Tesser2023Apr,Acciai2024Feb,Aguilar2024Sep}, namely distributions that can not be characterized by a temperature and an electrochemical potential, see Fig.~\ref{fig:setup} for a sketch of the considered system.  

On the one hand, the developed bounds set constraints on the achievable precision of a transport setup. Since we express these bounds in the form of measurable quantities, they provide trade-offs that can be used as guidelines to optimize a device with respect to noise, highlighting the important system parameters that are required to achieve high precision.
On the other hand, we anticipate that in particular, the TKUR, which is a tight bound in a large parameter range, can serve as an inference tool for estimating entropy production~\cite{Seifert2019Mar}.
While recent measurements have made first progress in measuring entropy~\cite{Hartman2018Nov,Child2022Nov,Pyurbeeva2022Dec,Piquard2023Nov}, entropy production still remains typically difficult to measure in quantum transport. Then, the TKUR-like bound allows for bounding the entropy production from below using measurements of charge currents and their fluctuations.

The remainder of this paper is organized as follows. We provide background on scattering theory for generic transport currents (such as charge, energy, or entropy) and their fluctuations in Sec.~\ref{sec:approach} and show how they connect to transport rates and activities in Sec.~\ref{sec:rates}. We establish a relation between entropy production in the case of broken time-reversal symmetry compared to the time-reversal symmetric case in Sec.~\ref{sec:TRS}, which will be used to establish general bounds for both cases. In Sec.~\ref{sec:precision_bounds}, we first show how the classical TUR, KUR, and TKUR are derived from scattering theory. Then, we extend all three to the quantum regime, both for the fermionic and the bosonic case, and express them in terms of measurable transport observables. Finally, we show in Sec.~\ref{sec:inference}, how an inverted TKUR-like bound can be used to infer the entropy production from charge transport measurements in generic transport setups. Details for the derivations of all bounds are given in the Appendices.

\section{Steady-state quantum transport}

\subsection{Scattering theory for generic currents and their fluctuations}\label{sec:approach}

We aim to describe coherent steady-state quantum transport in generic multi-terminal devices, as depicted in Fig.~\ref{fig:setup}, where terminals described by either fermionic or bosonic, possibly nonthermal contacts, are connected by a central scattering region by multi-channel leads. To this end, we restrict ourselves to studying systems with weak particle-particle interactions and utilize scattering theory, which allows for the treatment of strong coupling between contacts~\cite{Blanter2000Sep,Moskalets2011Sep}. The central region is described by a scattering matrix 
\begin{equation}
s(E)=\begin{bmatrix}
t_{11}(E) & t_{12}(E) & \dotsc \\
t_{21}(E) & t_{22}(E) & \dotsc \\
\vdots  & \vdots  & \ddots 
\end{bmatrix}
\end{equation}
where $t_{\alpha \beta}$ are energy-dependent  matrices of dimension $n_\alpha\times n_\beta$, which describe the transmission amplitudes {from the $n_\beta$ channels of the lead connected to reservoir $\beta$ into the $n_\alpha$ channels} of the lead connected to reservoir $\alpha$. 
Due to particle-current conservation, the scattering matrix must fulfill the unitarity condition,
\begin{equation}
    \sum_\beta t_{\alpha \beta}(E)t_{\alpha' \beta}^\dag(E) = \mathbb{I}_{\alpha}\delta_{\alpha\alpha'},
\end{equation}
where $ \mathbb{I}_{\alpha}$  is the $n_\alpha\times n_\alpha$ identity matrix. The transport of a generic quantity into reservoir $\alpha$ is described by the current
\begin{align}
    I^{(\nu)}_\alpha &=\langle\hat{I}_\alpha^{(\nu)}(t)\rangle 
    = \frac{1}{h} \int_0^\infty \!\!dE \sum_{\beta \neq \alpha} D_{\alpha \beta}(E) x^\nu_\alpha(E) (f_\beta(E)-f_\alpha(E)), \label{eq: current}
\end{align}
where we defined the transmission function as the sum over the channel-resolved probabilities of transmitting a particle from $\beta$ to $\alpha$
\begin{equation}\label{eq:Ddef}
    D_{\alpha \beta} (E)\equiv\text{tr}\left\{t_{\alpha \beta}(E) t_{\alpha \beta}^\dag(E)\right\} \leq n_\alpha,
\end{equation}
and $f_\alpha(E)$ is the average occupation of reservoir $\alpha$ at energy E.
Here, $\nu$ indicates the type of current going along with the real valued $x^\nu_\alpha(E)$ acting as a generalized charge associated with each current. For the particle current the label reads $\nu=N$ with $x^N_\alpha(E)=1$, for the energy current $\nu=E$ with $x^E_\alpha(E)=E$ and for the entropy current  $\nu=\Sigma$, it is $x^{\Sigma}_\alpha(E)=k_\mathrm{B} \log \left[(1\pm f_\alpha(E))/f_\alpha(E)\right]$~\cite{Deghi2020Jul,Acciai2024Feb}, where the upper and lower signs are for bosonic and fermionic systems respectively. Note that for thermal reservoirs the expression for the entropy current reduces to the Clausius relation, $I^{(\Sigma)}_\alpha=(I^{(E)}_\alpha-\mu_\alpha I^{(N)}_\alpha)/T_\alpha$ where $\mu_\alpha$ is the chemical potential and $T_\alpha$ is the temperature of the reservoir. From here on, for compactness, we suppress the energy arguments of transmission and occupation functions, and of the charge variable $x_\alpha^\nu(E)$, when they appear under the integration sign.

In this paper, we are interested in understanding the relation between an average current and its noise, which is defined through the zero-frequency autocorrelators~\cite{Blanter2000Sep}
\begin{equation}
    S_{\alpha\alpha}^{(\nu)}=\int_{-\infty}^{\infty} dt \langle \delta\hat{I}_\alpha^{(\nu)}(t)\delta\hat{I}_\alpha^{(\nu)}(0)\rangle,\label{eq:Sdef}
\end{equation}
where $\delta\hat{I}_\alpha^{(\nu)}(t)=\hat{I}_\alpha^{(\nu)}(t)-I_\alpha^{(\nu)}$.
Originally, the TUR, KUR, and TKUR were derived for classical Markovian systems~\cite{Barato2015Apr,Gingrich2016Mar,DiTerlizzi2018Dec,Vo2022Sep}, and therefore, to derive similar bounds for scattering theory, it is useful to split the fluctuations into a ``classical" and a ``quantum" part $S_{\alpha\alpha}^{(\nu)}=S_{\alpha\alpha}^{(\nu)\mathrm{cl}}+S_{\alpha\alpha}^{(\nu)\mathrm{qu}}$~\cite{Nazarov2009May}. The classical contribution
\begin{subequations}
\label{eq:S}
\begin{equation}\label{eq:Scl}
    S_{\alpha\alpha}^{(\nu)\mathrm{cl}}=\frac{1}{h}\int dE (x^{\nu}_{\alpha})^2 \biggr\{\sum_{\beta\neq\alpha}D_{\alpha \beta}( F^\pm_{\alpha \beta} + F^\pm_{\beta \alpha}) \biggr\} 
\end{equation}
is quadratic in the scattering matrix elements. We defined $F^\pm_{\alpha\beta}(E)\equiv f_\alpha(E)(1\pm f_\beta(E))$, where the upper sign is for bosonic systems and the lower one for fermionic systems, which henceforth will be the convention for the rest of the paper. The classical part of the fluctations can be understood as containing \textit{single-particle} effects, and it displays similar features to the steady-state noise found in classical Markovian systems~\cite{VanKampen2011Aug}. However, it still incorporates quantum effects both from the scattering matrix and from the factors $F^\pm_{\alpha\beta}(E)$, implementing the Pauli exclusion principle for fermionic systems and bunching for bosonic ones.

Quantum \textit{correlations} stemming from two-particle exchanges are, in contrast, included in the quantum contribution,
\begin{align}\label{eq:Squ}
    S^{(\nu)\text{qu}}_{\alpha \alpha}= \pm \frac{1}{h} \int dE&\sum_{\beta\neq\alpha}\sum_{\gamma\neq\alpha} \text{tr}\left\{t_{\alpha \beta} t_{\alpha \beta}^\dag t_{\alpha \gamma} t_{\alpha \gamma}^\dag\right\} 
    (x^\nu_\alpha(E))^2 (f_\beta-f_\alpha)(f_\gamma-f_\alpha)
\end{align}
\end{subequations}
which is quartic in the scattering matrix elements. Importantly, the quantum part of the fluctuations is positive in bosonic systems, increasing the noise, while it is negative in fermionic systems, decreasing the noise. Furthermore, the quantum part only contributes outside of equilibrium, namely for energies where the average occupations between reservoirs differ. With these expressions for average currents and their fluctuations, we move on to discussing how the noise relates to the activity of the system.

\subsection{Transport rates and activities}\label{sec:rates}

The classical contribution to the particle current noise, Eq.~\eqref{eq:Scl}, can be expressed in terms of single-particle transfer rates. These transfer rates out of and into reservoir $\alpha$ are 
\begin{subequations}\label{eq:rate-def}
\begin{eqnarray}
    \Gamma_{\alpha}^\to & = & \frac1h\sum_{\beta\neq\alpha}\int dE D_{\alpha\beta}F_{\alpha\beta}^\pm,\\
\Gamma_{\alpha}^\gets & = & \frac1h\sum_{\beta\neq\alpha}\int dE D_{\alpha\beta}F_{\beta\alpha}^\pm,
\end{eqnarray}    
\end{subequations}
respectively, and are valid also in the multi-channel case, see Eq.~\eqref{eq:Ddef}.
This allows us to write the classical particle-current fluctuations as the sum of transport rates~\cite{Nazarov2009May}
\begin{equation}\label{eq:activity}
S_{\alpha\alpha}^{(N)\mathrm{cl}}=\Gamma_{\alpha}^\to+\Gamma_{\alpha}^
\gets,
\end{equation}
and the average particle current as their difference
\begin{equation}
I_{\alpha}^{(N)}=\Gamma_{\alpha}^
\gets -\Gamma_{\alpha}^\to.
\end{equation}
 As a consequence, $S_{\alpha\alpha}^{(N)\mathrm{cl}}$ can be interpreted as the local activity $\mathcal{K}^\mathrm{cl}_\alpha$ of the reservoir $\alpha$ due to single-particle transfers~\cite{Palmqvist2024Oct} 
\begin{eqnarray}
\mathcal{K}^\text{cl}_\alpha & \equiv & \sum_{\beta\neq\alpha} \int dE D_{\alpha \beta }( F_{\alpha \beta}^{\pm}+F_{ \beta\alpha}^{\pm}) =  S_{\alpha \alpha}^{(N)\text{cl}},\label{eq:local_activity}\\
\mathcal{K}^\text{cl} &\equiv & \sum_{\alpha}  \mathcal{K}^\text{cl}_\alpha = \sum_\alpha S_{\alpha \alpha}^{(N)\text{cl}},\label{eq:global_activity}
\end{eqnarray}
where the second line hence defines the total activity due to single-particle transfers $\mathcal{K}^\mathrm{cl}$. It is important to note that the rates defined in Eqs.~\eqref{eq:rate-def} only count particle transfers between reservoirs, i.e., events where a particle is transmitted, contributing to transport. These rates do not count scattering events where, e.g., a particle is reflected back into another channel of the same lead~\cite{Blasi2024Nov}.
In the following, we refer to $\mathcal{K}^\text{cl}$ and $\mathcal{K}^\text{cl}_\alpha$ as the activity and the local activity, having in mind that this always refers to the ``classical" activity due to single-particle transfers.

\subsection{Entropy production and broken time-reversal symmetry}\label{sec:TRS}
For systems fulfilling time-reversal symmetry (TRS), the transmission functions are symmetric with respect to the reservoir indices $D_{\alpha \beta}(E)=D_{\beta \alpha}(E)$. However, in a general setting, TRS can be broken meaning that the transmissions instead fulfill the relation $D_{\alpha \beta}(E)=D^{\mathrm{tr}}_{ \beta\alpha}(E)$ where $D^{\mathrm{tr}}_{\alpha\beta}(E)$ is the transmission function of the time-reversed dual system~\cite{Blanter2000Sep,Moskalets2011Sep}. For example, in the transport of charged particles, if one applies a magnetic field $\mathbf{B}$ in the scattering region, the transmission satisfies $D_{\alpha \beta}(E;\mathbf{B}) =D^\mathrm{tr}_{ \beta\alpha}(E;\mathbf{B})=D_{ \beta\alpha}(E;-\mathbf{B})$ where the direction of the magnetic field is reversed in the dual system.

This has important consequences for transport quantities, in particular for the entropy production
\begin{equation}\label{eq: entropy prod}
    \sigma =\sum_\alpha I^{(\Sigma)}_\alpha.
\end{equation}
Indeed, previously derived thermodynamic bounds on fluctuations are often restricted to time-reversal symmetric systems or need to be modified in order to capture the case of broken time-reversal symmetry~\cite{Brandner2018Mar,Chun2019Apr,VanVu2019Sep,Lee2019Dec,Potanina2021Apr,Lee2021Nov,Pietzonka2022Apr,Brandner2025Feb}. Using the property of the transmission {functions} under time-reversal, we define the symmetrized entropy production,
\begin{align}
    \sigma^+ \equiv \frac{\sigma +\sigma^\mathrm{tr}}{2} = \sum_{\alpha}  \frac{I^{(\Sigma)}_{\alpha}+(I^{(\Sigma)}_{\alpha})^\mathrm{tr}}{2} 
    &= \frac{\kB}{2 h} \sum_{\alpha,\beta\neq\alpha}  \left\{ \int dE D_{\alpha\beta} \log\left[\frac{(1\pm f_\alpha)f_\beta}{(1\pm f_\beta)f_\alpha}\right] (f_\beta -f_\alpha) \right\} \nonumber \\
    &= \frac{\kB}{2 h} \sum_{\alpha,\beta\neq\alpha}  \left\{ \int dE D_{\alpha \beta} \log\left[\frac{F_{\beta \alpha}^\pm}{F_{\alpha \beta}^\pm}\right] \left(F_{\beta \alpha}^\pm-F_{\alpha \beta }^\pm\right) \right\}, \label{eq: step1 TUR TS}
\end{align}
which is identical to the entropy production if there is TRS, e.g., in the case of vanishing magnetic fields in charge transport, $\sigma^+=\sigma= \sum_{\alpha}  I^{(\Sigma)}_{\alpha}(\mathbf{B}=0)$. This expression for the entropy production in time-reversal symmetric systems was recently used in Ref.~\cite{Brandner2025Feb}. 

In order to relate the $\sigma^+$ to $\sigma$, we derive in Appendix~\ref{app: sigma plus bound deriv} the following generally valid inequality,
\begin{equation}\label{eq: sigma plus bound}
    \sigma^+ \leq{\lambda} \sigma ,  
\end{equation}
where we defined the asymmetry in the transmissions,
\begin{equation}
\label{eq: def asymmetry}
    \lambda \equiv \max_{E,\alpha,\beta} \frac{D_{\alpha \beta}(E) +D_{\beta\alpha }(E)}{2D_{\alpha \beta}(E) }
\end{equation}
quantifying to what extent TRS is broken. Bound~\eqref{eq: sigma plus bound} is useful for extending {the uncertainty relations} derived for systems with TRS to systems without TRS as long as the transport is not unidirectional (meaning as long as $D_{\alpha\beta}(E)\neq0$ if $D_{\beta\alpha}(E)\neq0$) and it is the key ingredient in generalizing the bounds developed in this paper to systems with broken TRS.

\section{Precision bounds}\label{sec:precision_bounds}

In this section, we present bounds on the precision of multi-terminal bosonic and fermionic scattering systems. These bounds have the character of a kinetic uncertainty relation (KUR), which constrains precision by the activity, or of a thermodynamic uncertainty relation (TUR), which constrains precision by the entropy production.
Importantly, we also derive combined bounds, the so-called thermodynamic kinetic uncertainty relations (TKURs). We start with a quantum-transport formulation of these bounds in the limit of weak transmissions or weak biases, corresponding largely to the classical regime, in Sec.~\ref{sec:Cl_bounds}, and subsequently derive bounds in the full quantum limit in Secs.~\ref{sec:quantum_bosons} and \ref{sec:quantum_fermions}.

\subsection{``Classical" limit of weak transmission or small bias}\label{sec:Cl_bounds}

To derive trade-off relations for precision, we begin by studying the classical fluctuations. We thereby find trade-off relations that are analogous to existing classical bounds. For quantum transport, these relations are particularly relevant for small biases or in the weak transmission limit, i.e., for $\sum_\beta D_{\alpha \beta}(E)|f_\beta(E)-f_\alpha(E)|\ll1$ $\forall E$, when the classical part of the fluctuations, Eq.~\eqref{eq:Scl}, is the dominant term in the noise.  

\subsubsection{Kinetic uncertainty relation (KUR)}\label{sec:Cl_KUR}

In order to derive a classical KUR from quantum transport, we start from the simple inequality for the distribution functions~\cite{Acciai2024Feb,Palmqvist2024Oct} occurring in the integrands of the noise contributions
\begin{equation}\label{eq:f_inequality}
    D_{\alpha\beta}(E)\left|f_\beta(E)-f_\alpha(E)\right|=D_{\alpha\beta}(E)\left|F_{\beta \alpha}^\pm(E) -F_{ \alpha\beta}^\pm(E) \right|\leq D_{\alpha\beta}(E)\left(F_{\beta \alpha}^\pm(E) +F_{ \alpha\beta}^\pm(E)\right).
\end{equation}
Exploiting furthermore Jensen's inequality, see Appendix~\ref{app:derive_bounds} for details, we find a bound on the precision $\mathcal{P}_\alpha^{(\nu)\text{cl}}$ with respect to the classical noise of a current $I_\alpha^{(\nu)}$ in contact $\alpha$,
\begin{equation}\label{eq: cl KUR}
     \mathcal{P}_\alpha^{(\nu)\text{cl}} \equiv \frac{\left( I^{(\nu)}_\alpha \right)^2 }{S^{(\nu)\text{cl}}_{\alpha \alpha}} \leq  S^{(N)\text{cl}}_{\alpha \alpha} =\mathcal{K}^\mathrm{cl}_\alpha .
\end{equation}
This is a \textit{local} version of the KUR, which in the case of single-channel leads coincides with the result found in Ref.~\cite{Palmqvist2024Oct}.
Here, no assumptions about the scattering matrix or average occupations were made, apart from the fact that bosonic distributions fulfill $f_\alpha(E)\geq 0$ and fermionic distributions fulfill the Pauli exclusion principle and hence $1\geq f_\alpha(E) \geq0$. 

To make the connection with the kinetic uncertainty relation, the right-hand side of the bound, namely the classical contribution to the particle current noise, is identified with the \textit{local} activity, see Eq.~(\ref{eq:local_activity}), namely with respect to one reservoir $\alpha$, only.

\subsubsection{Thermodynamic uncertainty relation (TUR)}\label{sec:Cl_TUR}

Thermodynamic uncertainty relations in thermal fermionic systems described by scattering theory have previously been established in Ref.~\cite{Brandner2018Mar, Potanina2021Apr,Brandner2025Feb}. Furthermore, the TUR was investigated and shown to be valid for two thermal bosonic reservoirs connected by an arbitrary harmonic coupling at a temperature bias in Ref.~\cite{Saryal2019Oct}. 

We show here that the classical TUR holds for both bosonic and fermionic systems, possibly even with nonthermal reservoirs. Starting from the fact that the logarithmic mean is smaller than the arithmetic mean \cite{Kwon2024Dec}
\begin{equation}\label{eq: step 1 TUR}
    \frac{F_{\beta \alpha}^{\pm}(E)-F_{\alpha \beta }^{\pm}(E)}{\log[F^\pm_{\beta\alpha }(E)]-\log[F^\pm_{\alpha \beta}(E)]} \leq \frac{F_{\beta \alpha}^{\pm}(E)+F_{\alpha \beta }^{\pm}(E)}{2},   
\end{equation} 
and applying Jensen's inequality, it follows that
\begin{equation}
     \sum_\alpha \mathcal{P}^{(\nu)\mathrm{cl}}_\alpha = \sum_\alpha \frac{\left( I^{(\nu)}_\alpha \right)^2}{S_{\alpha \alpha}^{(\nu)\text{cl}}} \leq \frac{{\sigma}^+}{\kB} \leq \frac{\lambda {\sigma}}{\kB}.\label{eq:cl_TUR}
\end{equation}
Here, the first inequality contains the symmetrized entropy production, while the second one holds for the full entropy production even in systems where time-reversal symmetry is broken. 
 Importantly, we use an extension to systems with broken time-reversal symmetry, valid for both fermionic \textit{and} bosonic systems that is different from the extension used for fermionic systems in Refs.~\cite{Brandner2018Mar,Potanina2021Apr,Brandner2025Feb}, where a numerical factor of $1/(0.8956...)$ was inserted in front of $\sigma$. We note that the use of the factor $\lambda$, see~\eqref{eq: sigma plus bound}, for fermionic systems yields a  tighter bound for systems with weak time-reversal symmetry breaking, while the use of the numerical factor of~\cite{Brandner2018Mar,Potanina2021Apr,Brandner2025Feb} yields a tighter bound for systems with strong time-reversal symmetry breaking. 
For a detailed derivation of~\eqref{eq:cl_TUR}, see Appendix~\ref{app:derive_bounds}.

\subsubsection{Thermodynamic kinetic uncertainty relation (TKUR)}\label{sec:Cl_TKUR}

The TUR provides a tight bound on precision close to equilibrium, whereas the KUR saturates far from equilibrium. 
These bounds hence play complementary roles, and they serve as performance limits in their respective regimes. 
However, it is desirable to have a bound that remains tight in a broader range of biases. 
Such a bound, referred to as the unified thermodynamic kinetic uncertainty relation (TKUR), was introduced in Ref.~\cite{Vo2022Sep} for classical Markovian systems satisfying local detailed balance. 
Here, we extend the TKUR to systems described by scattering theory. We therefore start from the equality~\cite{Vo2022Sep}
\begin{equation}\label{eq:Vo_equality}
    \tanh\left[\frac{1}{2} \log [a/b] \right] = \frac{a-b}{b+a}
\end{equation}
and choose $a=F^\pm_{\beta \alpha}(E)$ and $b=F^\pm_{\alpha\beta}(E)$, namely the products of distribution functions entering the rates and hence the classical activity. As explained in detail in  Appendix~\ref{app: deriv TKUR}, we then integrate over energies and sum over contacts to obtain the desired thermodynamic and transport observables. Applying Jensen's inequality, this leads us to the  bound 
\begin{equation}\label{eq: TKUR cl TRS}
    \sum_\alpha \frac{\left( I^{(\nu)}_\alpha \right)^2}{S_{\alpha \alpha}^{(\nu)\text{cl}}} \leq \frac{\sigma^+}{\kB} \xi\left[\frac{\sigma^+}{\kB \mathcal{K}^\mathrm{cl} }\right]
\end{equation}
where we introduced the function
\begin{equation}\label{eq:xi}
    \xi[x] \equiv \frac{x}{\Omega^2[x]},
\end{equation}
and defined $\Omega[x]$ as the inverse function of $x \tanh{x}$. The precision is hence bounded by a function of the classical activity and the symmetrized entropy production.

In order to extend this bound such that it contains the full entropy production, independently of the properties of the system under time-reversal, we resort to~\eqref{eq: sigma plus bound}. Indeed, multiplying $\sigma^+$ with the factor $\lambda$ both in the numerator of the right-hand side of~\eqref{eq: TKUR cl TRS} and in the argument of the function $\Omega[x]$ is compatible with the TKUR, thanks to
\begin{equation}\label{eq:Omega_derivative}
    \frac{d}{dx} \frac{x^2}{y \Omega^2[x/y]} \geq 0
\end{equation}
for $x,y,\geq0$. We can hence  extend the TKUR bound to systems without TRS
\begin{equation}\label{eq: TKUR cl}
    \sum_\alpha \mathcal{P}^{(\nu)\mathrm{cl}}_\alpha \leq \frac{\lambda\sigma}{\kB} \xi\left[\frac{\lambda\sigma}{\kB \mathcal{K}^\mathrm{cl} }\right].
\end{equation}
The bound~\eqref{eq: TKUR cl} is a key result of this paper, extending the TKUR to the classical precision of a current in scattering theory even for systems with broken time-reversal symmetry. 
Close to equilibrium or in the weak transmission limit, the classical contribution to the noise approximately equals the measurable fluctuations, allowing the TKUR~\eqref{eq: TKUR cl} to serve as a powerful inference tool for estimating entropy production, see Sec.~\ref{sec:inference} for more detailed results and discussion of inference. 

In the extreme case of unidirectional transport, i.e., $D_{\alpha \beta}(E)>0$ but $D_{\beta \alpha}(E)=0$, the factor $\lambda$ diverges, $\lambda\rightarrow\infty$. This causes the bound~\eqref{eq: TKUR cl} to reduce to a KUR for the total activity,
\begin{equation}
     \sum_\alpha \mathcal{P}^{(\nu)\mathrm{cl}}_\alpha \leq \mathcal{K^\mathrm{cl}},
\end{equation}
since $\lim_{x\rightarrow\infty} \frac{x^2}{\Omega^2[x]}=1$ \cite{Vo2022Sep}. This means that due to the factor $\lambda$, entropy production does not have a significant weight in the TKUR~\eqref{eq: TKUR cl}, when time-reversal symmetry breaking is significant.

Equations~\eqref{eq: TKUR cl TRS} and \eqref{eq: TKUR cl} involve a sum over the precisions in all contacts as well as the full activity due to all current rates. In situations in which the precision of a current in one specific contact is of interest, a bound on this single precision is desirable. By using~\eqref{eq:Omega_derivative} as detailed in Appendix~\ref{app: deriv TKUR}, an analogous local TKUR can be derived, which is given by
\begin{equation}\label{eq:localTKUR_cl}
    \mathcal{P}^{(\nu)
    \mathrm{cl}}_\alpha  \leq \frac{\lambda\sigma}{{2}\kB} \xi\left[\frac{\lambda\sigma}{{2}\kB \mathcal{K}_\alpha^\mathrm{cl} }\right].
\end{equation}
It has the important feature that the right-hand side only contains the \textit{local} activity. However, $\sigma$ is still the full entropy production: as a consequence, the local bound~\eqref{eq:localTKUR_cl} is less tight than the full classical TKUR~\eqref{eq: TKUR cl}, when entropy production takes place to a significant extent without involving reservoir $\alpha$.
\begin{figure}[t]
    \centering
    \begin{minipage}{0.49\textwidth}
        \centering
        \includegraphics[width=\linewidth]{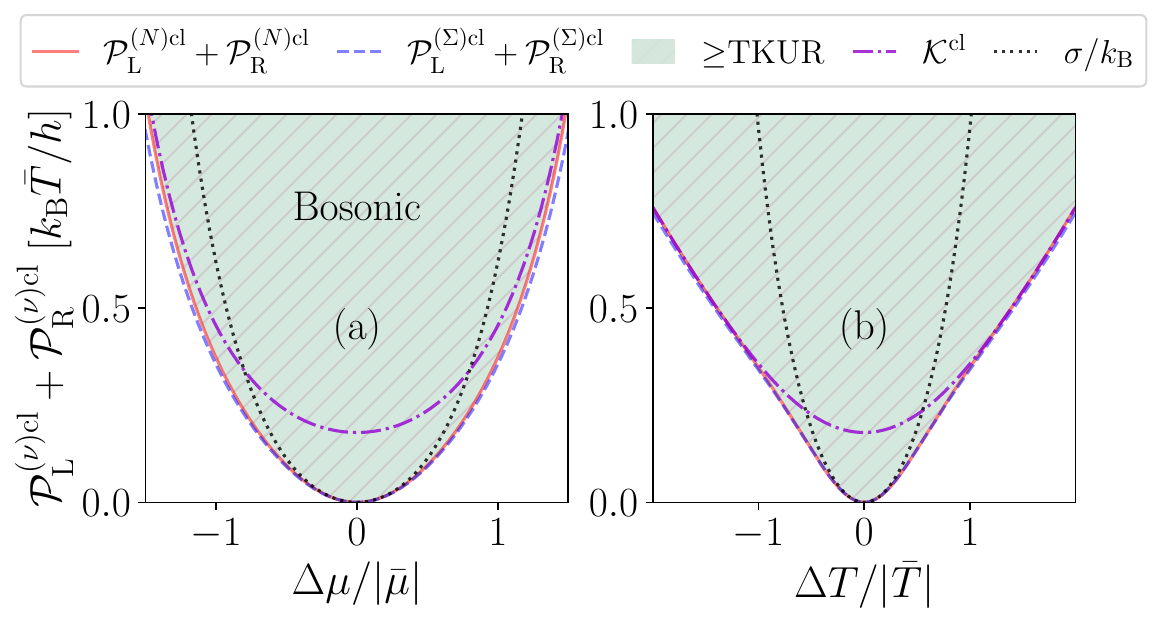}
    \end{minipage}
    \hfill
    \begin{minipage}{0.49\textwidth}
        \centering
        \includegraphics[width=\linewidth]{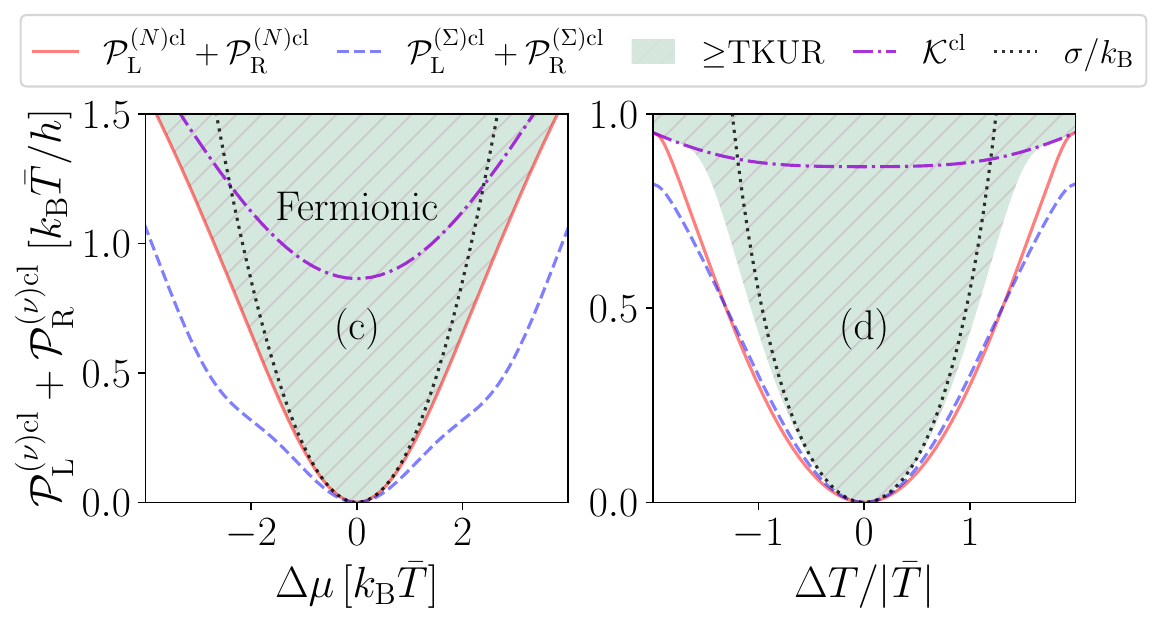}
    \end{minipage}
    \caption{KUR~\eqref{eq: cl KUR}, TUR~\eqref{eq:cl_TUR} and TKUR~\eqref{eq: TKUR cl} in a thermal two-terminal system with a boxcar transmission. Bounds set by entropy production (black dotted lines),  total single particle transfer activity (purple dash-dotted lines), and TKUR (green filled surfaces) constrain the summed precision functions for particle current (red lines) and for entropy current (blue dashed lines). Panels~(a,c) show functions of chemical potential bias with $\Delta T=0$ and panels~(b,d) show functions of temperature bias with $\Delta\mu=0$. Panels~(a,b) displays the bounds in a bosonic system with $\Bar{\mu}=-3\,\kB\Bar{T}$, $D=1$, $E_0=0.1\,\kB\Bar{T}$ and $b=3\,\kB\Bar{T}$. Panels~(b,d) displays the bounds in a fermionic system with $\Bar{\mu}=0$, $D=0.5$, $E_0=0.1\,\kB\Bar{T}$ and $b=3\,\kB\Bar{T}$. Here what is meant by $\geq\mathrm{TKUR}$ are all values greater than the function ${\lambda\sigma} \xi\left[{\lambda\sigma}/(\kB \mathcal{K}^\mathrm{cl} )\right]/{\kB}$, see~\eqref{eq: TKUR cl}.}  
    \label{fig:cl_bounds}
\end{figure}

To illustrate the behaviour of the classical bounds~\eqref{eq: cl KUR}, \eqref{eq:cl_TUR}, and \eqref{eq: TKUR cl}, we calculate the precisions, activities and entropy production in a two-terminal system, with thermal reservoirs at temperatures $T_\mathrm{L} = \Bar{T} +\Delta T/2$, $T_\mathrm{R} = \Bar{T} -\Delta T/2$ and chemical potentials  $\mu_\mathrm{L} = \Bar{\mu} +\Delta \mu/2$, $\mu_\mathrm{R} = \Bar{\mu} -\Delta \mu/2$. Since we consider thermal reservoirs, entropy currents and fluctuations are obtained using the Clausius relation,  $x^\Sigma_\alpha = (E-\mu_\alpha)/T_\alpha$. The contacts are connected by a single-channel boxcar transmission $D_\mathrm{LR}(E) =D \left\{\theta(E-E_0)-\theta(E-E_0-b)\right\}$. 
Here, $\theta(E)$ is the Heaviside function, $b$ is the width of the transmission window, and $E_0$ is the onset of the region with constant transmission $D\neq 0$ as illustrated in Fig.~\ref{fig:setup}. Transmission functions approaching this boxcar shape can be achieved using chains of quantum dots~\cite{Whitney2015Mar} in electronic conductors, wavelength-selective mirrors for optical systems, or multipole Purcell filters in circuit QED~\cite{Yan2023Sep}. 

In Fig.~\ref{fig:cl_bounds}, we show the sum of classical precisions $\mathcal{P}^{(\nu)\mathrm{cl}}_\mathrm{L}+\mathcal{P}^{(\nu)\mathrm{cl}}_\mathrm{R}$ for both particle current (red lines) and the entropy current (blue dashed lines) for bosonic systems in panels~(a,b) and for fermionic systems in panels~(c,d). Both precision functions, $\mathcal{P}^{(N)\mathrm{cl}}_\mathrm{L}+\mathcal{P}^{(N)\mathrm{cl}}_\mathrm{R}$ and $\mathcal{P}^{(\Sigma)\mathrm{cl}}_\mathrm{L}+\mathcal{P}^{(\Sigma)\mathrm{cl}}_\mathrm{R}$, display similar behavior becoming larger as the applied bias increases. The reduction of the entropy-current precision compared to the particle-current precision, visible as ``shoulders" in panel~(c), can be explained by the fact that all of the entropy can be produced in only one of the reservoirs in a fermionic system, meaning that one of the precisions goes to zero at a given potential bias.

Figure~\ref{fig:cl_bounds} furthermore shows the bounds on the precisions set by the TUR~\eqref{eq:cl_TUR} given by the total entropy production (dotted black lines), by the activity (purple dashdotted lines) constraining precision via the KUR~\eqref{eq: cl KUR}, and the TKUR of~\eqref{eq: TKUR cl} as green filled surface, indicating all values greater than the function $\xi\left[{\lambda\sigma}/(\kB \mathcal{K}^\mathrm{cl} )\right]/{\kB}$. One clearly observes that the TUR constrains the precision well close to equilibrium, but as the bias is increased, it fails to provide a tight bound. On the other hand, the KUR fails to provide a tight bound at equilibrium as there is still thermal noise present, but it is instead tight far from equilibrium. Specifically when one of the transition rates is dominating $|I^{(N)}_L| \approx |\Gamma^{\rightleftharpoons}_L|$,  the KUR becomes tight. The TKUR interpolates between TUR and KUR.
Indeed the TKUR is shown to coincide with $\sigma$ at small biases and with the total activity $\mathcal{K}^\mathrm{cl}$ at large biases~\cite{Vo2022Sep} [This is even more clearly seen in panels~(a,e) of Fig.~\ref{fig:Quantum precision bounds}]. Moreover, the TKUR can be seen to provide a tighter bound than either the TUR or the KUR in an intermediate bias regime, allowing us to make a stricter statement about the achievable precision in these regimes.

This section shows how to recover and extend bounds analogous to the ones for classical Markovian systems, by focusing on the classical part of the noise and its impact on the precision. We are, however, interested in general out-of-equilibrium situations without restricting ourselves to the weak transmission or close-to-equilibrium regime. It is hence not enough to address the classical noise, but bounds on the full measurable fluctuations are required. Importantly, the quantum part of the noise has very different features for bosons and fermions, see also Ref.~\cite{Palmqvist2024Oct}, and we hence address these different situations separately.

\subsection{Full quantum limit - bosonic systems}\label{sec:quantum_bosons}

\subsubsection{Simple extension to full precision}
For bosons, the full fluctuations are larger than the classical part, since the quantum contribution, Eq.~\eqref{eq:Squ}, is always positive. This means that the full precision is suppressed with respect to the precision of ``classical" systems, where the quantum contribution to the noise can be neglected,
\begin{equation}
    \mathcal{P}^{(\nu)}_{\alpha } \equiv \frac{\left( I^{(\nu)}_{\alpha} \right)^2}{S^{(\nu)}_{\alpha\alpha}} \leq \mathcal{P}^{(\nu)\text{cl}}_{\alpha }.
\end{equation}
Thus, both the TUR and KUR can be extended to the full fluctuations,
\begin{align}
    \mathcal{P}^{(\nu)}_{\alpha}  & \leq S^{(N)\mathrm{cl}}_{\alpha \alpha} \leq S^{(N)}_{\alpha \alpha}, \label{eq:KUR_full_simple}\\
   \sum_\alpha \mathcal{P}^{(\nu)}_{ \alpha} &\leq \frac{\lambda\sigma}{\kB}.  \label{eq:TUR_full_simple}
\end{align}
We extend the bosonic TKUR in the same fashion, using
\begin{equation}\label{eq: Omega dy}
    \frac{d}{dy} \frac{x^2}{y \Omega^2[x/y]} \geq0,
\end{equation}
for $x,y\geq0$, and the fact that the full fluctuations are increased when the quantum part of the noise is added, to get
\begin{align}\label{eq:TKUR_full_simple}
    \sum_\alpha \mathcal{P}^{(\nu)}_{\alpha} &\leq \frac{\lambda\sigma}{\kB} \xi\left[ \frac{1}{\kB} \frac{\lambda\sigma}{\left(\sum_\alpha{S}^{(N)}_{\alpha\alpha}\right)} \right],\\
    \mathcal{P}^{(\nu)}_{\alpha} &\leq \frac{\lambda\sigma}{2\kB} \xi\left[ \frac{\lambda\sigma}{2{\kB}{S}^{(N)}_{\alpha\alpha}} \right].
\end{align}
Note that $\sum_\alpha{S}^{(N)}_{\alpha\alpha}$ and $S^{(N)}_{\alpha \alpha}$ no longer represent the actual activity of the system. They however bound the total and local activities $\mathcal{K}^\text{cl}$ and $\mathcal{K}^\mathrm{cl}_\alpha$ from above.

While these bounds provide simple trade-off relations for precision, they are only tight when the quantum noise can be neglected. A tighter bound under nonequilibrium and large transmission conditions requires additional bounds on the quantum fluctuations.

\subsubsection{Bound on quantum fluctuations}
In this section, we provide a bound for the quantum contribution to the noise, only. In particular, such a bound can subsequently be used to estimate the classical activity from the full noise. For bosonic systems, the single-channel case was addressed in Ref.~\cite{Palmqvist2024Oct} using the Cauchy-Schwarz inequality. The full derivation of the extension to the multi-channel case is shown in Appendix~\ref{app: bosonic q bounds}; here we present the main results. We denote the indicator function of the spectral current as
\begin{equation}
\zeta_\alpha^\nu (E)\equiv
    \begin{cases}
         1 \quad \text{if} \quad E\in\text{supp}\{x_\alpha^{\nu}(E)\sum_\beta D_{\alpha\beta}(E)\left(f_\beta(E)-f_\alpha(E)\right)\},  \\
         0 \quad \text{otherwise}
    \end{cases}
\end{equation}
which we use to define the bandwidth of transport 
\begin{equation}
    B_\alpha^\nu \equiv \int dE \zeta_\alpha^\nu (E).
\end{equation}
We prove in Appendix~\ref{app: bosonic q bounds} that the precision in any current is limited by the number of active channels $n_\alpha$ and the bandwidth of transport,
\begin{equation}\label{eq: Boson Q bound}
    \frac{n_\alpha B^{\nu}_\alpha}{ h} \geq \frac{\left({I^{(\nu)}_{\alpha}}\right)^2 }{ S^{(\nu)\text{qu}}_{\alpha \alpha}} \geq \frac{\left({I^{(\nu)}_{\alpha}}\right)^2 }{ S^{(\nu)}_{\alpha \alpha}}. 
\end{equation}
This bound, which generalizes the single-channel result of Ref.~\cite{Palmqvist2024Oct}, shows that the precision in any bosonic current is inherently limited by the bandwidth and the amount of available channels, due to bunching effects. Specifically, a small bandwidth and a small number of channels imply fewer distinct states available to particles, thereby enhancing bunching. The inequality~\eqref{eq: Boson Q bound} is tight when the transmission of different channels is equal and when the bandwidth is smaller than the energy scale over which $\sum_\alpha D_{\alpha \beta}(E) x^\nu_\alpha(E) |f_\beta(E) -f_\alpha(E)|$ changes, reflecting the fact that the transport of a quantity is maximally spread out between different channels and available energy states, avoiding bunching.

\subsubsection{TUR, KUR-, and TKUR-like bounds in the quantum regime}

Since the bandwidth constrains the quantum part of the noise and hence the full precision of any current $I_{\alpha}^{(\nu)}$ in the form given by~\eqref{eq: Boson Q bound}, it is now possible to extend the KUR, TUR, and TKUR from Sec.~\ref{sec:Cl_bounds} to hold for a function of the \textit{full precision}. Concretely, we find for bosonic quantum systems subject to bunching 
\begin{align}
    \phantom{\sum_\alpha}\mathcal{W}^\nu_\alpha \left[\mathcal{P}^{(\nu)}_{\alpha} \right]& \leq {S}^{(N)\mathrm{cl}}_{\alpha\alpha},\label{eq: boson q KUR}\\
    \label{eq: boson q TUR}
    \sum_\alpha \mathcal{W}^\nu_\alpha \left[\mathcal{P}^{(\nu)}_{\alpha} \right]&\leq\frac{\lambda\sigma}{\kB}, \\
    \sum_\alpha \mathcal{W}^\nu_\alpha \left[\mathcal{P}^{(\nu)}_{\alpha} \right]&\leq \frac{\lambda\sigma}{\kB} \xi\left[\frac{\lambda\sigma}{\kB \mathcal{K}^\mathrm{cl} }\right] \label{eq: boson q TKUR},
\end{align}
where we introduced the function
\begin{equation}\label{eq:def_W}
    \mathcal{W}_\alpha^\nu[x] \equiv \frac{x}{1-\frac{h}{B_\alpha^\nu}x} \geq x\ .
\end{equation}
using~\eqref{eq: Boson Q bound}. By taking into account bunching effects in the precision, these bounds are more constraining than the simple extensions to the quantum regime provided in~\eqref{eq:KUR_full_simple}, \eqref{eq:TUR_full_simple} and \eqref{eq:TKUR_full_simple}, which do not use the function $\mathcal{W}^\nu_\alpha[x]$. 

The bounds~\eqref{eq: boson q KUR} and \eqref{eq: boson q TKUR} contain the activity due to single-particle transfers, $\mathcal{K}^\mathrm{cl}$ given in Eq.~\eqref{eq:global_activity}, a quantity which in general is not directly observable in a transport experiment. Our next goal is hence to establish bounds, which are not only valid in the quantum regime, but are also expressed in terms of observable quantities only. In order to reach this goal, we need to find an estimate for the actual activity in terms of transport observables.

Using the inequality~\eqref{eq: Boson Q bound} for the quantum part, we construct an upper bound on the classical part of the fluctuations
\begin{equation}\label{eq: boson S tilde}
   {S}^{(\nu)\text{cl}}_{\alpha\alpha}\leq   {S}^{(\nu)}_{\alpha\alpha}-\frac{ h}{n_\alpha B^{\nu}_\alpha} \left({I^{(\nu)}_{\alpha}}\right)^2. 
\end{equation}
This inequality allows us to estimate the activity due to single-particle transfers in and out of a reservoir, from the particle current, its fluctuations, the bandwidth, and the number of active channels. 
We use this bound on the classical part of the noise to estimate the classical activity by measurable quantities, namely
\begin{equation}\label{eq:activity_estimate}
    \mathcal{K}^{\mathrm{cl}}_{\alpha}\leq 
     {S}^{(N)}_{\alpha\alpha}-\frac{ h}{n_\alpha B^{N}_\alpha} \left({I^{(N)}_{\alpha}}\right)^2\equiv \widetilde{\mathcal{K}}_{\alpha, \mathrm{bos}}  .
\end{equation}
These estimates are based on the bound~\eqref{eq: Boson Q bound}, and they therefore approach equality when transport is maximally spread out between different channels and available energy states, avoiding bunching. Furthermore to estimate the total activity we define {$\sum_\alpha\widetilde{\mathcal{K}}_{\alpha, \mathrm{bos}} \equiv \widetilde{ \mathcal{K}}_\mathrm{bos}$}.

Using~\eqref{eq:activity_estimate},  relating the single-particle activity to transport observables in the quantum regime together with~\eqref{eq: Omega dy}, we are able to write tight bounds on the full precision in terms of the experimentally accessible quantities,
\begin{align}
    \label{eq: full boson q KUR}
\phantom{\sum_\alpha}\mathcal{W}^\nu_\alpha \left[\mathcal{P}^{(\nu)}_{\alpha} \right]&\leq
     {S}^{(N)\mathrm{cl}}_{\alpha\alpha} \leq \widetilde{\mathcal{K}}_{\alpha,\text{bos}} \\
    \label{eq: boson full q TKUR}
    \sum_\alpha \mathcal{W}^\nu_\alpha \left[\mathcal{P}^{(\nu)}_{\alpha}\right]&\leq \frac{\lambda\sigma}{\kB} \xi\left[ \frac{1}{\kB} \frac{\lambda\sigma}{\widetilde{\mathcal{K}}_\text{bos}} \right] .
\end{align}

It is also possible to extend the \textit{local} TKUR in an analogous fashion
\begin{equation}\label{eq:full_local_bosonTKUR}
    \mathcal{W}^\nu_\alpha \left[\mathcal{P}^{(\nu)}_\alpha\right] \leq  \frac{\lambda \sigma}{{2}\kB} \xi\left[ \frac{\lambda\sigma}{{2}\kB\mathcal{K}^\text{cl}_\alpha} \right]
     \leq  \frac{\lambda \sigma}{{2}\kB} \xi\left[\frac{\lambda\sigma}{2 \kB\widetilde{\mathcal{K}}_{\alpha,\mathrm{bos}}} \right].
\end{equation}

\begin{figure*}[t]
    \centering
    \includegraphics[width=6.6in]{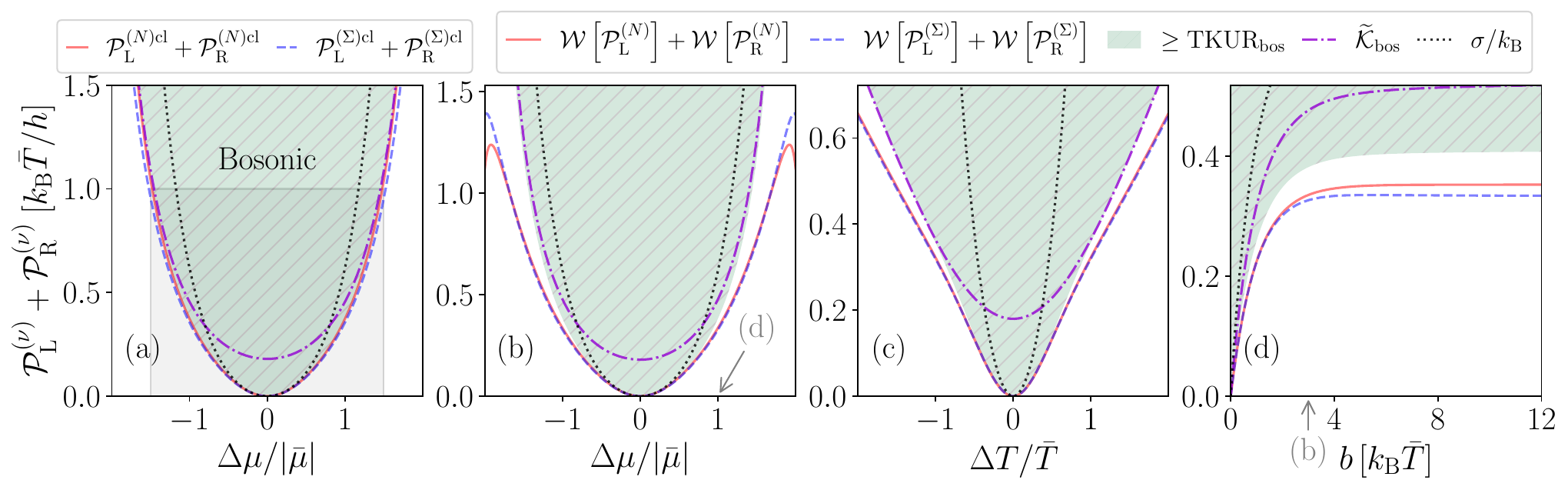}
    \par
    \includegraphics[width=6.6in]{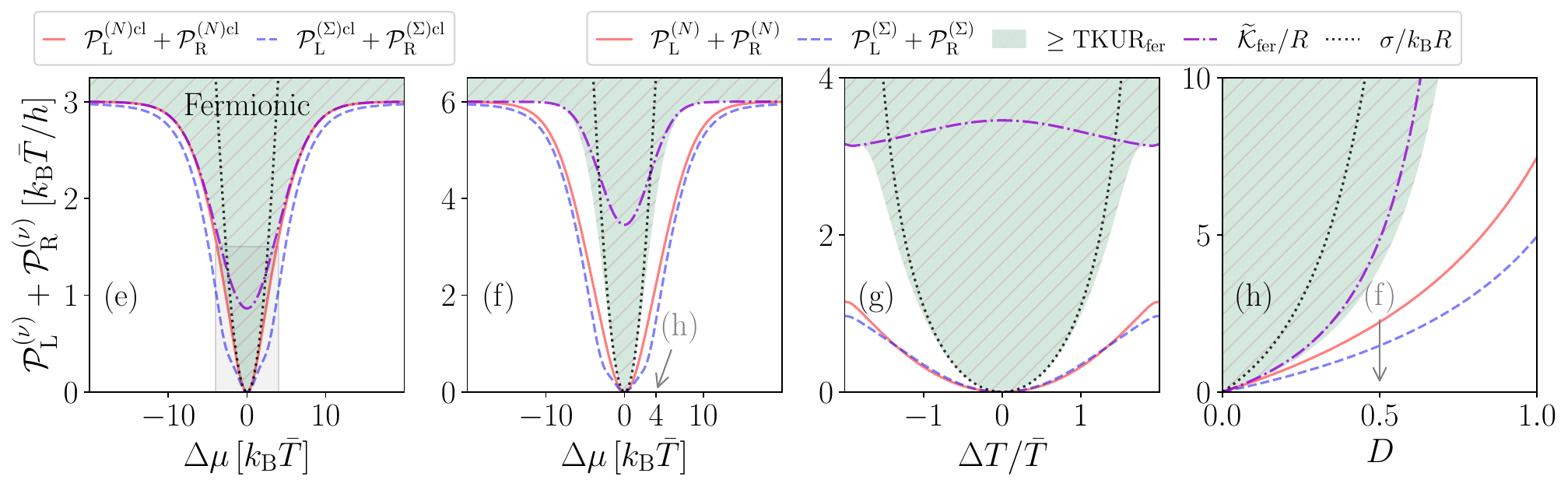}
     \caption{TUR-, KUR- TKUR-like bounds in a thermal two-terminal system with boxcar transmission. Summed precisions of the particle currents are displayed in red lines and of the entropy currents in the blue dashed lines. Panels~(a,e) display the classical bounds where the gray area shows the axis limits of panels (a,c) of Fig.~\ref{fig:cl_bounds}. Panels~(b,f) display the full bounds---purple dash-dotted line for the KURL for bosons~\eqref{eq: full boson q KUR} and fermions~\eqref{eq: fermion full q KUR}, dotted black line for the TUR for bosons~\eqref{eq: boson q TUR} and fermions~\eqref{eq: fermion q TUR}, and filled green surface for the TKURL for bosons~\eqref{eq: boson full q TKUR} and fermions~\eqref{eq: fermion full q TKUR} displaying all values greater than respective upper bound---as function of chemical potential bias with $\Delta T=0$. Panels~(c,g) show the bounds as a function of temperature bias with $\Delta\mu=0$. 
    For bosonic systems, panels~(a-d) use $\Bar{\mu}=-3\,\kB\Bar{T}$ and panels~(a-c) have a boxcar transmission parametrized by $D=1$, $E_0=0.1\,\kB\Bar{T}$ and $b=3\,\kB\Bar{T}$. Panel~(d) displays the bounds as function of width of the transmission $b$ at $\Delta\mu=\kB\bar{T}$ using $D=1$ and $E_0=0.1\,\kB\Bar{T}$.
    For fermionic systems, panels~(e-h) use $\Bar{\mu}=0$ and the boxcar transmission of panels~(e-g) are parametrized by  $D=0.5$, $E_0=0.1\,\kB\Bar{T}$ and $b=3\,\kB\Bar{T}$. Panel~(f) displays the bounds as a function of transmission $D$ and uses $\Delta\mu=4\kB\bar{T}$, $\Delta T= 0$, $E_0=0.1\,\kB\Bar{T}$ and $b=3\,\kB\Bar{T}$.
 } 
    \label{fig:Quantum precision bounds}
\end{figure*}

To show the result for the bosonic bounds~\eqref{eq: boson q TUR}, \eqref{eq: full boson q KUR},  and \eqref{eq: boson full q TKUR}, in the full quantum limit, we again consider two thermal reservoirs connected by a single-channel boxcar transmission with bandwidth $B^\nu_\mathrm{L}=B^\nu_\mathrm{R} =b$. 
We here introduce $\mathcal{W}[x]=x/(1-xh/b)$, which is similar to Eq.~\eqref{eq:def_W} but for a symmetric scatterer. The bounds on functions of precisions $\mathcal{W}[ \mathcal{P}^{(\nu)}_\mathrm{R}]+\mathcal{W}[ \mathcal{P}^{(\nu)}_\mathrm{L}]$  are shown for the particle current (red lines) and entropy current (blue dashed lines) in panels~(b-d) of Fig.~\ref{fig:Quantum precision bounds} as functions of chemical potential bias, temperature bias and bandwidth $b$. 
These functions of precision are constrained by the entropy production (dotted black line) through the TUR~\eqref{eq: boson q TUR}, with the bound being tight at equilibrium. 
Away from equilibrium, the KUR-like (KURL) relation~\eqref{eq: full boson q KUR} provides a better bound set by the estimate of the activity $\widetilde{\mathcal{K}}_\mathrm{bos}$ (dash-dotted purple lines). 
However, the KURL is found not to be generally tight far from equilibrium, contrary to the classical bound, as shown in panels~(b) and (c) compared to the classical limit shown in panel~(a) of Fig.~\ref{fig:Quantum precision bounds}. The reasons for this are the following: when one of the chemical potentials approaches zero, the Bose-Einstein distribution diverges at zero energy, causing the precision to decrease due to bunching.
At the same time, the bound~\eqref{eq: Boson Q bound} fails to provide a good estimate of the classical noise since the average occupations vary strongly in the transmission window. Instead, the KURL provides a tight bound far from equilibrium when the average occupations vary slowly compared to the energy scale set by $b$. This is shown in panel~(d) of Fig.~\ref{fig:Quantum precision bounds}, where for small bandwidths the bound approaches the precision functions. The bound set by the TKUR-like (TKURL) relation~\eqref{eq: boson full q TKUR} is displayed as a green filled area, indicating all values greater than the function ${\lambda\sigma}\xi[ {\lambda\sigma}/(\kB{\widetilde{\mathcal{K}}_\text{bos}})]/{\kB}$. Utilizing the total entropy production and the estimate of the activity $\widetilde{\mathcal{K}}_\mathrm{bos}$, the TKURL is shown to interpolate between the bounds set by the TUR and KURL. It thereby provides a substantially tighter limit on precision, especially for larger bandwidths and intermediate biases. Also, the TKURL gets as tight as the classical bound for small bandwidths.

\subsection{Full quantum limit -- Fermionic systems}\label{sec:quantum_fermions}

In fermionic systems, the quantum contribution to noise is negative, such that the bounds established in the classical limit no longer hold~\cite{Brandner2018Mar,Potanina2021Apr,Acciai2024Feb,Palmqvist2024Oct}. Indeed, in fermionic systems, the quantum contribution  \textit{increases} the precision. Therefore, a simple extension of the bounds, as done in Sec.~\ref{sec:quantum_bosons} for bosonic systems, is not possible here. Instead, we need to find bounds for fermionic systems that are able to capture the increase in precision in the quantum regime.

\subsubsection{Bound on quantum fluctuations}

As in the bosonic case, we start by estimating the quantum contributions to the fluctuations. Since in fermionic systems, transport can be noiseless in the absence of reflection due to antibunching, it can be expected that the impact of the quantum part of the noise depends on the extent to which reflection occurs.
Indeed, using  $(f_\beta(E)-f_\alpha(E))\in[-1,1]$ for fermionic average occupations, as well as the unitarity of the scattering matrix, we show that the integrand of the quantum contribution to the noise is bounded by the classical one up to a reflection-factor
\begin{align}
   &\sum_{\beta,\gamma\neq\alpha} \text{tr}\left\{t_{\alpha \beta} t_{\alpha \beta}^\dag t_{\alpha \gamma} t_{\alpha \gamma}^\dag\right\} (x^\nu_\alpha)^2 (f_\beta-f_\alpha)(f_\gamma-f_\alpha) 
   \leq\sum_{\gamma\neq\alpha} \text{tr}\left\{\left(\mathbbm{1}_\alpha - t_{\alpha\alpha}t_{\alpha\alpha}^\dagger\right) t_{\alpha \gamma} t_{\alpha \gamma}^\dag\right\} (x^\nu_\alpha)^2 \left(F_{\gamma\alpha}+F_{\alpha\gamma}\right),
\end{align}
see Appendix~\ref{app: fermionic q bounds} for details. For the full integrated noise expressions, this implies
\begin{equation}\label{eq: fermionic q bound}
    \frac{S^{(\nu)}_{\alpha \alpha}}{R_\alpha} \geq S^{(\nu)\text{cl}}_{\alpha \alpha},
\end{equation}
where we define the minimum reflection probability $R_\alpha \equiv \min_i \inf_{E \in A}R_{\alpha,i}(E)$ and $R_{\alpha,i}(E)$ are the eigenvalues of $t_{\alpha \alpha}(E)t_{\alpha \alpha}^\dag(E)$ labeled by $i\in\{1,2,...,n_\alpha\}$. Here $A$ denotes the support of the spectral current, i.e., of the integrand of Eq.~\eqref{eq: current}. Applying~\eqref{eq: fermionic q bound} to the particle-current noise, we hence show that the full fermionic noise bounds the classical noise contribution, which in turn represents the single-particle activity. In the single-channel limit,~\eqref{eq: fermionic q bound} coincides with the results obtained in Ref.~\cite{Palmqvist2024Oct}.

\subsubsection{TUR, KUR-, and TKUR-like bounds in the quantum regime}

The bound of~\eqref{eq: fermionic q bound} allows us to extend the classical bounds of Sec.~\ref{sec:Cl_bounds} to constraints on the full precision of fermionic transport
\begin{align}
    R_\alpha\mathcal{P}^{(\nu)}_{\alpha} & \leq S_{\alpha \alpha}^{(N)\mathrm{cl}} ,\\
    \label{eq: fermion q TUR}
     \sum_\alpha R_\alpha \mathcal{P}^{(\nu)}_{\alpha}& \leq \frac{\lambda\sigma}{\kB}, \\
    \sum_\alpha R_\alpha \mathcal{P}^{(\nu)}_{\alpha}& \leq \frac{\lambda\sigma}{\kB} \xi\left[ \frac{1}{\kB} \frac{\lambda\sigma}{\mathcal{K}^\mathrm{cl}} \right] ,\\
    R_\alpha \mathcal{P}^{(\nu)}_{\alpha}& \leq \frac{\lambda\sigma}{{2}\kB} \xi\left[ \frac{\lambda\sigma}{{2} \kB S^{(N)\text{cl}}_{\alpha \alpha}} \right].
\end{align}
These relations bound the full precision by entropy production and by the \textit{classical}, single-particle activity. In order to find similar bounds for measurable quantities, we again use~\eqref{eq: fermionic q bound} and establish a bound on the activity using the full measurable noise
\begin{equation}
    \mathcal{K}^\mathrm{cl} \leq\sum \frac{1}{R_\alpha} S_{\alpha \alpha}^{(N)} \equiv \widetilde{\mathcal{K}}_\mathrm{fer}  .
\end{equation}
This, in turn, allows us to write KUR-like (KURL) and TKUR-like (TKURL) bounds in terms of the full fluctuations
\begin{align}
    \label{eq: fermion full q KUR}
    R_\alpha\mathcal{P}^{(\nu)}_{\alpha} & \leq\frac{1}{R_\alpha}S_{\alpha \alpha}^{(N)}, \\
    \label{eq: fermion full q TKUR}
   \sum_\alpha R_\alpha \mathcal{P}^{(\nu)}_{\alpha}& \leq\frac{\lambda\sigma}{\kB} \xi\left[ \frac{1}{\kB} \frac{\lambda\sigma}{\mathcal{K}_\text{fer}} \right], \\
   \label{eq: local fermion full q TKUR}
   R_\alpha \mathcal{P}^{(\nu)}_{\alpha}& \leq\frac{\lambda\sigma}{{2}\kB} \xi\left[ \frac{R_\alpha}{{2}\kB} \frac{\lambda\sigma}{S_{\alpha \alpha}^{(N)}} \right].
\end{align}
To show the behaviour of our fermionic bounds on the full noise, we again consider the thermal two-terminal system connected by a single-channel boxcar transmission, meaning that $R\equiv R_\mathrm{L}=R_\mathrm{R}=1-D$. 
The sums of precisions $\mathcal{P}^{(\nu)}_\mathrm{L}+\mathcal{P}^{(\nu)}_\mathrm{R}$ are shown for the particle current (red lines) and entropy current (blue dashed lines) in panels~(f-h) of Fig.~\ref{fig:Quantum precision bounds} as functions of chemical potential bias, temperature bias and transmission probability $D$. 
For the precision functions, the main difference from the classical case is that they get larger since the full noise is smaller than the classical noise due to anti-bunching, which is made evident by comparing panel~(e) to panel~(f) of Fig.~\ref{fig:Quantum precision bounds}. Furthermore, the important role of the reflection probability in achieving precision is displayed in panel~(h), where the precision grows as $D$ increases.

The bound on the sum of precisions set by the TUR~\eqref{eq: fermion q TUR} $\sigma /\kB R$ (dotted black lines) can again be seen to work best close to equilibrium. 
The constraint set by the KURL~\eqref{eq: fermion full q KUR} $\widetilde{\mathcal{K}}_\mathrm{fer} /R$ (dashdotted purple line) is shown to work well at large chemical potential biases. However, the KURL is not in general tight when only a temperature bias is applied. Also in the fermionic full quantum regime, the TKURL~\eqref{eq: fermion full q TKUR} shown as a green filled area displaying all values greater than the function ${\lambda\sigma}\xi[ {\lambda\sigma}/(\kB{\widetilde{\mathcal{K}}_\text{fer}})]/(R{\kB})$ is shown to interpolate between the TUR and KURL with an intermediate bias regime when it provides a better bound than either of the two. Panel~(h) clearly shows the important role played by the transmission probability in how tight the bounds are; all upper bounds diverge in the limit of full transmission, thereby providing only trivial bounds on the precisions when the reflection is zero. This is in particular visible in the case we show, as we here chose $\Delta\mu=4\kB \bar{T}$, where indeed the bounds are not expected to be tight, see panel~(f).

\subsection{Bounds on entropy production --- Thermodynamic inference}\label{sec:inference}

It is, in general, a difficult task to measure entropy production directly in an experiment~\cite{Hartman2018Nov,Child2022Nov,Pyurbeeva2022Dec,Piquard2023Nov}. This becomes even more complicated if the contacts are nonthermal and entropy production can no longer be directly related to heat flows via the Clausius relation~\cite{Pekola2021Oct}. 
A tool to access entropy production is ``inference", where other observables, known to bound the entropy production, are measured instead~\cite{Seifert2019Mar}, providing a nontrivial bound. Inference is, however, only truly predictive when the underlying bounds are reasonably tight. 

For multi-terminal quantum transport devices, as studied here, this is provided by the TKUR-like bounds~\eqref{eq: boson full q TKUR}, \eqref{eq:full_local_bosonTKUR}, \eqref{eq: fermion full q TKUR} and \eqref{eq: local fermion full q TKUR}. In order to use these bounds for inference, we hence need to find closed expressions setting a bound on $\sigma$ or $\sigma^+$~\cite{Vo2022Sep}.
For compactness, we now introduce a short notation for different functions of the precision and related activity estimates, see Table~\ref{tab:boson_fermion_shorthand}.
\begin{table}[b]
    \centering
    \renewcommand{\arraystretch}{2.1} 
    \begin{tabular}{|c|c|c|c|c|}
        \hline
        \multicolumn{5}{|c|}{\textbf{Bosonic Bounds}} \\ \hline
         $\boldsymbol{\mathcal{P}}$ & $\sum_\alpha \mathcal{W^\nu_\alpha\left[\mathcal{P}^{(\nu)}_\alpha \right] }$ & $\sum_\alpha \mathcal{P}^{(\nu)}_\alpha$ & ${2}\mathcal{W^\nu_\alpha\left[\mathcal{P}^{(\nu)}_\alpha \right] }$ &${2}\mathcal{P}^{(\nu)}_\alpha$ \\ 
        \hline
        $\boldsymbol{\mathcal{K}}$ & $\mathcal{K}^\mathrm{cl},\widetilde{\mathcal{K}}_\mathrm{bos}$ & $\sum_\alpha S^{(N)}_{\alpha \alpha}$ & ${2}\mathcal{K}_\alpha^\mathrm{cl},{2}\widetilde{\mathcal{K}}_{\alpha,\mathrm{bos}}$ &${2}S^{(N)}_{\alpha \alpha}$\\ 
        \hline
        \multicolumn{5}{|c|}{\textbf{Fermionic Bounds}} \\ \hline
         $\boldsymbol{\mathcal{P}}$ & $\sum_\alpha R_\alpha \mathcal{P}^{(\nu)}_\alpha $ & ${2}R_\alpha \mathcal{P}^{(\nu)}_\alpha$ & \multicolumn{2}{c|}{} \\ 
        \hline
        $\boldsymbol{\mathcal{K}}$ & $\mathcal{K}^\mathrm{cl},\widetilde{\mathcal{K}}_\mathrm{fer}$ & ${2}\mathcal{K}_\alpha^\mathrm{cl},{2}S^{(N)}_{\alpha \alpha} /R_\alpha$ & \multicolumn{2}{c|}{} \\ 
        \hline
    \end{tabular}
    \caption{Notation for bosonic and fermionic bounds. When $\boldsymbol{\mathcal{P}_\alpha}$ or $\boldsymbol{\mathcal{K}_\alpha}$ is used in the text, this denotes one element of the sums defining the short notation, e.g. $\boldsymbol{\mathcal{P}_\alpha}$ could symbolize $R_\alpha \mathcal{P}^{(\nu)}_\alpha$ when referring to the fermionic bounds~(\ref{eq: fermion q TUR},\ref{eq: fermion full q KUR},\ref{eq: fermion full q TKUR}) or $\mathcal{W}^\nu_\alpha\left[\mathcal{P}^{(\nu)}_\alpha\right]$ for the bosonic bounds~(\ref{eq: boson q TUR},\ref{eq: full boson q KUR},\ref{eq: boson full q TKUR}).}
    \label{tab:boson_fermion_shorthand}
\end{table}
\begin{figure}[h!]
    \centering
    \includegraphics[width=1\columnwidth]{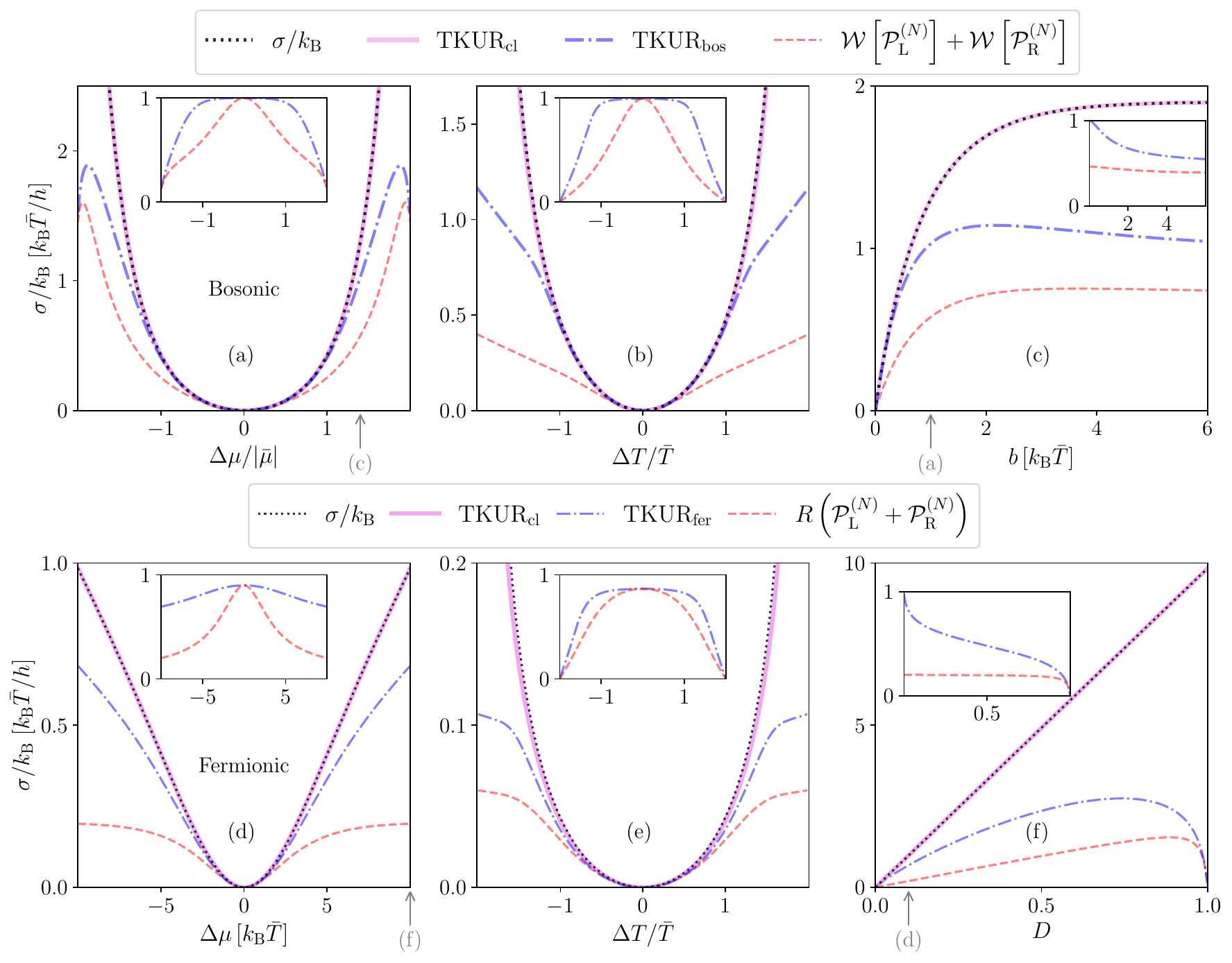}
    \caption{TURs~(\ref{eq: boson q TUR},\ref{eq: fermion q TUR}) and TKURLs~\eqref{eq:inference_bound} applied as inference tools for estimating the entropy production $\sigma$ (dotted black lines) from particle currents and fluctuations in thermal bosonic (upper row) and fermionic (lower row) two-terminal systems. Panels~(a,d) show bounds as a function of $\Delta\mu$ with $\Delta T=0$ and panels~(b,e) show the bounds as a function of $\Delta T$ with $\Delta\mu=0$. 
    The red dashed lines display the bound set by the TURs~(\ref{eq: boson q TUR},\ref{eq: fermion q TUR}). The bounds set by the TKURLs~\eqref{eq:inference_bound} using the full noises with quantum corrections are displayed as blue dash-dotted lines. The pink line displays the classical TKUR~\eqref{eq:inference_bound} using the classical noise and activity.
    Panels~(a,b,c) show the bounds for a bosonic system and have $\Bar{\mu}=-3\,\kB\Bar{T}$ and panels~(a,b) have a boxcar transmission parametrized by $D=1$, $E_0=0.1\,\kB\Bar{T}$ and $b=\kB\Bar{T}$. Panel~(c) shows the bounds as function of the transmission width $b$ at $\Delta \mu=1.4\kB\bar{T}$ with $D=1$ and $E_0=0.1\,\kB\Bar{T}$.
    Panels~(d,e,f) show the bounds for a fermionic system and have $\Bar{\mu}=0$ and panel~(d) has a boxcar transmission parametrized by $D=0.1$, $E_0=0.1\,\kB\Bar{T}$ and $b=\kB\Bar{T}$ and panel~(e) has $D=0.1$, $E_0=\kB\Bar{T}$ and $b=\kB\Bar{T}$. Panel~(f) shows the bounds as function of the transmission probability $D$ at $\Delta \mu=10\kB\bar{T}$ with $E_0=0.1\,\kB\Bar{T}$ and $b=\kB\Bar{T}$. Insets show the lower bounds divided by $\sigma/\kB$, displaying the saturation of our bounds, i.e. $\kB\boldsymbol{\mathcal{P}}/ (\lambda \sigma)\leq 1$ for the TURs and ${\kB}\log\left[{(\sqrt{\boldsymbol{\mathcal{K}}}+\sqrt{\boldsymbol{\mathcal{P}}})}/{(\sqrt{\boldsymbol{\mathcal{K}}}-\sqrt{\boldsymbol{\mathcal{P}}})} \right]\sqrt{\boldsymbol{\mathcal{K}}\boldsymbol{\mathcal{P}}}/( \lambda \sigma) \leq 1$ for the TKURLs.
    }
    \label{fig:Inference}
\end{figure}
With these short notations, we start from 
\begin{equation}
    \boldsymbol{\mathcal{P}} \leq \frac{(\sigma\lambda)^2}{\kB^2 \boldsymbol{\mathcal{K}}} \Omega^{-2}\left[ \frac{\sigma\lambda}{\kB \boldsymbol{\mathcal{K}}}\right]
\end{equation}
representing the bounds~\eqref{eq: boson full q TKUR}, \eqref{eq:full_local_bosonTKUR}, \eqref{eq: fermion full q TKUR} and~\eqref{eq: local fermion full q TKUR}.
Using that $\Omega[x]$ is the inverse function of $ x \tanh x$ and that it is strictly increasing for $x>0$, we find 
\begin{equation}
    \sqrt{\frac{\boldsymbol{\mathcal{P}}}{\boldsymbol{\mathcal{K}}}} \leq \tanh \left[\frac{\lambda\sigma/\kB}{\sqrt{\boldsymbol{\mathcal{K}} \boldsymbol{\mathcal{P}}}}\right].
\end{equation}
Exploiting $\boldsymbol{\mathcal{P}}/{\boldsymbol{\mathcal{K}}}\leq1$ via all the KUR-like bounds, the inverse of the $\tanh$-function can be expressed as $\tanh^{-1}{x} =\frac{1}{2}\log\left[(1+x)/(1-x)\right]$ \cite{Gradshteyn2014}, (see also Eq.~\eqref{eq:Vo_equality}) and we find a lower bound on entropy production
\begin{equation}\label{eq:inference_bound}
     \frac{1}{2}\log\left[\frac{\sqrt{\boldsymbol{\mathcal{K}}}+\sqrt{\boldsymbol{\mathcal{P}}}}{\sqrt{\boldsymbol{\mathcal{K}}}-\sqrt{\boldsymbol{\mathcal{P}}}} \right]\sqrt{\boldsymbol{\mathcal{K}}\boldsymbol{\mathcal{P}}} \leq \frac{\lambda\sigma }{\kB}.
\end{equation}
We recall that, here, $\boldsymbol{\mathcal{K}}$ is the short notation for any of the activity-like quantities and $\boldsymbol{\mathcal{P}}$ the short notation for the related local precisions or sums over them. This means that~\eqref{eq:inference_bound} provides two sets of bounds to infer the entropy production in fermionic or bosonic quantum transport settings by local (or global) measurements of charge currents and their fluctuations. 

To illustrate the application of our bounds as inference tools we return to the two-terminal system with thermal reservoirs connected by single-channel boxcar transmission and compare the bounds set by the TKURLs~\eqref{eq:inference_bound} and the TURs~(\ref{eq: boson q TUR},\ref{eq: fermion q TUR}). For bosonic systems, we show this in panels~(a,b,c) of Fig.~\ref{fig:Inference} for varying chemical potential bias, temperature bias, and bandwidth $b$. The total entropy production (dotted black lines) are bounded from below by $\mathcal{W}[\mathcal{P}^{(N)}_\mathrm{L}]+\mathcal{W}[\mathcal{P}^{(N)}_\mathrm{R}]$ via the TUR~\eqref{eq: boson q TUR} (dashed red lines). The TUR yields a good estimate close to equilibrium but is less predictive away from equilibrium. 
We show two versions of the TKURLs~\eqref{eq:inference_bound} for inference, one using single-particle transfer activities with $\boldsymbol{\mathcal{K}} = {\mathcal{K}}^\mathrm{cl}$ and $\boldsymbol{\mathcal{P}}=(I^{(N)}_\mathrm{L})^2 /\mathcal{K}^\mathrm{cl}_\mathrm{L} + (I^{(N)}_\mathrm{R})^2 /\mathcal{K}^\mathrm{cl}_\mathrm{R}$ (pink lines) and one using the full noise with $\boldsymbol{\mathcal{K}} =\widetilde{\mathcal{K}}_\mathrm{bos}$ and  $\boldsymbol{\mathcal{P}}=\mathcal{W}[\mathcal{P}^{(N)}_\mathrm{L}]+\mathcal{W}[\mathcal{P}^{(N)}_\mathrm{R}]$ (dash-dotted blue lines). All bounds work relatively well for small biases, with differences becoming clear away from equilibrium. The classical TKURL provides the most predictive results for all considered regimes, while the full TKURL is less predictive for large biases. However, for small transmission windows, the full TKURL also becomes similarly predictive as the classical TKURL even far from equilibrium and for full transparency $D=1$, where the quantum part of the noise has a pronounced effect. This can be seen in panel~(c). It means that if one is able to achieve a narrow filter, the bosonic TKURL provides a good estimate of $\sigma$ using the full, measurable noise. Moreover, we emphasize that the TKURL using the full noise is a substantially tighter bound than the TUR utilizing the exact same quantities for various considered parameter values.

We show the corresponding results for fermionic systems in panels~(d,e,f) of Fig.~\ref{fig:Inference} for varying chemical potential bias, temperature bias, and transmission probability $D$. The total entropy production (dotted black lines) is bounded by the summed precisions (red lines) via the TUR~\eqref{eq: fermion q TUR}, which can be seen to provide a relatively good estimate of $\sigma$ around equilibrium. It, however, fails to do so as the applied bias increases. In particular, when the applied potential bias approaches the width of the transmission, the precision stops increasing due to the Pauli exclusion principle while $\sigma$ continues growing, which is seen in panel~(d).
We again show two versions of TKURLs~\eqref{eq:inference_bound}, one using only the particle currents and the classical activity (pink lines) and one using the particle currents and their full noise with $\boldsymbol{\mathcal{K}} = \widetilde{\mathcal{K}}_\mathrm{fer}$ and $\boldsymbol{\mathcal{P}}=R(\mathcal{P}^{(N)}_\mathrm{L}+\mathcal{P}^{(N)}_\mathrm{R} )$ (blue dash-dotted lines). Once more, the three bounds behave similarly at equilibrium, but show significant differences for larger bias. In particular, the classical TKURL provides a very good estimate of $\sigma$ in a wide range of parameter values. 
The difference between the TKURL using the full noise and classical noise becomes larger as the bias is increased, especially in the high transparency limit, made visible in panel~(f). Comparing with panel~(f) of Fig.~\ref{fig:Quantum precision bounds} one might expect the bound on $\sigma$ of panel~(d) of Fig.~\ref{fig:Inference} to be more predictive as the bound on precision of Fig.~\ref{fig:Quantum precision bounds} is rather tight. This seemingly different behaviour is explained by the fact that the inverse function of $x^2/\Omega^2[x]$ diverges as $x\rightarrow1$, meaning that if the bound on the full precision is slightly worse than the classical bound, this leads to a larger discrepancy once the bound is inverted in~\eqref{eq:inference_bound}. However, the full TKURL still provides a substantially better constraint on $\sigma$ than the TUR, despite them being constructed from the same measurable quantities. This means that one can find a better estimate of $\sigma$ by simply inserting the particle currents, noise, and minimum reflection probabilities in the function of~\eqref{eq:inference_bound}, instead of combining them in the precision used in the TUR~\eqref{eq: fermion q TUR}. We also point out the fact that the bounds using the full noise do not saturate at unity for zero bias, with them instead taking the value $R$ as displayed in the insets of the fermionic plots.

Finally, we conclude this section by presenting a refined version of the inference bound~\eqref{eq:inference_bound}. As shown in Appendix~\ref{app: deriv TKUR}, a similar---yet generally tighter---bound can be derived by following the same approach as for the TKURs, 
\begin{equation}\label{eq:inference_bound2}
    \sum_\alpha\frac{1}{2}\log\left[\frac{\sqrt{\boldsymbol{\mathcal{K}}_\alpha}+\sqrt{\boldsymbol{\mathcal{P}}_\alpha} }{\sqrt{\boldsymbol{\mathcal{K}}_\alpha}-\sqrt{\boldsymbol{\mathcal{P}}_\alpha}}\right]\sqrt{\boldsymbol{\mathcal{K}}_\alpha\boldsymbol{\mathcal{P}}_\alpha}\leq\frac{\lambda \sigma }{\kB},
\end{equation}
where $\boldsymbol{\mathcal{P}}_\alpha$ and $\boldsymbol{\mathcal{K}}_\alpha$ denote individual terms within the sums used in the shorthand notation. This improvement is possible, since precisions and activities of single reservoirs, $\boldsymbol{\mathcal{P}}_\alpha$ and $\boldsymbol{\mathcal{K}}_\alpha$, were used, such that Jensen's inequality was used fewer times in the derivation; note that this would not have been possible for the derivation of the TKURL~\eqref{eq: TKUR cl}, since it has the form of a bound on the precision. 
As with~\eqref{eq:inference_bound}, the bound~\eqref{eq:inference_bound2} provides two sets of constraints for bosonic and fermionic systems. In general,~\eqref{eq:inference_bound2} offers a more predictive lower bound on entropy production in \textit{multi}-terminal systems. However, for the two-terminal cases considered in Fig.~\ref{fig:Inference}, the bounds~\eqref{eq:inference_bound} and \eqref{eq:inference_bound2} coincide due to particle-current conservation. 

Furthermore, we can gain additional intuition for the fundamental underlying principle behind these TKUR-like bounds by studying~\eqref{eq:inference_bound2}. By inserting $\boldsymbol{\mathcal{P}}_\alpha= \mathcal{P}^{(N)\mathrm{cl}}_\alpha$ and $\boldsymbol{\mathcal{K}}_\alpha=\mathcal{K}^\mathrm{cl}_\alpha$ into~\eqref{eq:inference_bound2} we can rewrite the bound in terms of single-particle transfer rates
\begin{equation}
    \sum_\alpha\frac{1}{2}\log\left[\frac{{\Gamma^\leftarrow_\alpha}}{\Gamma^\rightarrow_\alpha}\right](\Gamma^\leftarrow_\alpha -\Gamma^\rightarrow_\alpha)\leq\frac{\lambda \sigma }{\kB}.
\end{equation}
Comparing the left-hand side of this expression with the expression of $\sigma^+$ as seen in~Eq.~\eqref{eq: step1 TUR TS}, we note that \eqref{eq:inference_bound2} contains ``coarse-grained" versions of the \textit{spectral} particle-transfer rates determining the symmetrized entropy production.
This connection highlights that one can, to some extent, interpret the TKURs as a coarse-graining of the entropy production in time-reversal symmetric systems where \eqref{eq: sigma plus bound} is used to extend the bound to systems with broken TRS. By measuring classical local activity and precision of the particle current (experimentally accessible for small biases or weak transmission, see Sec.~\ref{sec:Cl_bounds}), one can hence estimate these coarse-grained rates. Moreover, we note that for the purpose of inferring entropy production, the actual activities due to single-particle transfers $\mathcal{K}^\mathrm{cl}_\alpha$ would result in better estimates than the full noise would. In a realistic quantum transport experiment, access to the full precision is, however, more straightforward. To summarize, the TKURs can be interpreted as providing a lower bound on entropy production by estimating coarse-grained rates of transfers in and out of a reservoir in terms of currents, noise, and estimates of classical activities.
\section{Conclusions}

We have provided bounds on the precision in quantum transport for currents of generic type---be it charge, energy, or entropy currents. These bounds take the form of TURs or KUR-like and TKUR-like constraints. The latter, predictive over a large range of nonequilibrium conditions, had not been considered for quantum transport settings until now. Here we provide versions of all three bounds that hold in the quantum regime, both for fermionic and bosonic systems, and even if time-reversal symmetry is broken or the contacts are characterized by nonthermal distributions.

Importantly, in the quantum regime, we express all bounds in terms of experimentally accessible transport observables (charge currents and their fluctuations, as well as transport bandwidth or reflection coefficients). This yields intuitive insights into which physical properties limit the achievable precision in a setup. Hence, we anticipate that our results will be useful as guidelines for the optimization of quantum transport devices, with the goal of reaching outputs with high precision. Furthermore, we show how the TKUR can be used to infer the entropy production, which is typically hard to access, from the measurement of charge transport alone, even if there are nonthermal contacts present in the setup. 

Our bounds, being derived from scattering theory, have the advantage of taking fully into account quantum effects at the cost of treating many-body effects up to the mean-field level only. An interesting question would be to find out how strong interactions impact the bounds developed here. An open challenge is furthermore how to transfer the concept of activity---of relevance for both the KUR and the TKUR---to systems in the quantum regime. While our work can give estimates on how quantum observables relate to the classical activity, a generally valid definition of a quantum activity is still outstanding. Furthermore, a way to obtain the single-particle transfer rates or better estimates of them in terms of experimentally accessible quantities would be interesting for improving the bounds used for inference of entropy production.

\section{Data availability}
Relevant data and derivations are available in the paper, and plots are obtained from straightforward calculation of expressions.
\begin{acknowledgments}
    We would like to thank Bruno Bertin Johannet for providing useful comments on our manuscript.
    Furthermore, we gratefully acknowledge funding from the European Research Council (ERC) under the European Union’s Horizon Europe research and innovation program (101088169/NanoRecycle) (D.P. and J.S.) and from the Knut and Alice Wallenberg foundation via the fellowship program (L.T. and J.S.). 
\end{acknowledgments}

\appendix

\section{Lower bound on entropy production in systems with broken time-reversal symmetry}\label{app: sigma plus bound deriv}

In order to find a lower bound on entropy production for systems with broken time-reversal symmetry (TRS), we begin by writing the total entropy production $\sigma$ as a sum of positive terms. We first define the function~\cite{Brandner2025Feb}
\begin{equation}
    \rho[x] \equiv -x \log[x] \pm (1 \pm x) \log[1 \pm x].
\end{equation}
where we use the upper signs to describe bosonic systems and the lower signs to describe fermionic systems. The parameter $x$ will be replaced by distribution functions, which hence have the restriction that $x\geq0$ for bosons and $1\geq x\geq 0$ for fermionic systems. Therefore, $\rho[x]$ is concave, allowing us to define the positive terms
\begin{equation}
    \rho_{\alpha \beta}(E) \equiv \rho[f_\alpha(E)]-\rho[f_\beta(E)] -\rho'[f_\alpha(E)](f_\alpha(E) -f_\beta(E)) \geq 0.
\end{equation}
Using the unitarity of the scattering matrix, we are now able to express the entropy production as
\begin{equation}
    \sigma =\sum_\alpha I^{(\Sigma)}_\alpha = \frac{\kB}{h}\int dE \sum_{\alpha\beta} D_{\alpha \beta} \rho_{\alpha \beta}
\end{equation}
where each term inside the sum is positive. This and similar expressions for the fermionic entropy production have previously been used in Refs.~\cite{Nenciu2007Mar,Brandner2018Mar,Potanina2021Apr,Brandner2025Feb} to treat $\sigma$ in systems where TRS is broken. Furthermore, we notice that
\begin{align}
    &\frac{\kB}{h}\int dE \sum_{\alpha\beta} D_{\alpha \beta} \rho_{\alpha \beta} \frac{D_{\alpha \beta} +D_{ \beta \alpha}}{D_{\alpha \beta}} = \frac{\kB}{h}\int dE \sum_{\alpha\beta} D_{\alpha \beta}(\rho_{\alpha \beta} +\rho_{\beta \alpha }) = 2 \sigma^+,
\end{align}
where we used the property $D_{\alpha \beta}(E) =D^\mathrm{tr}_{ \beta\alpha}(E)$ and renamed summation indices. Hence, by defining the asymmetry of the couplings 
\begin{equation}
    \lambda \equiv \max_{E,\alpha,\beta} \frac{D_{\alpha \beta}(E) +D_{\beta\alpha }(E)}{2D_{\alpha \beta}(E) }
\end{equation}
see also Eq.~\eqref{eq: def asymmetry} in the main text, we find the bound
\begin{equation}
    \lambda \sigma \geq \sigma^+.
\end{equation}

\section{Detailed derivation of classical thermodynamic and kinetic uncertainty relations}\label{app:derive_bounds}
In Ref.~\cite{Palmqvist2024Oct}, we derived Kinetic Uncertainty Relation-like (KURL) bounds for the single-channel case of scattering theory. This was achieved by minimizing a quadratic form containing an average arbitrary current and its classical fluctuations as well as the classical fluctuations of the particle current. Here we extend this relation to the multi-channel case using a different derivation. We begin with the relation~\cite{Acciai2024Feb},
\begin{equation}
    |f_\beta(E)-f_\alpha(E)|=|F_{\beta \alpha}^\pm(E) -F_{ \alpha\beta}^\pm(E)| \leq F_{\beta \alpha}^\pm (E)+F_{ \alpha\beta}^\pm(E).
\end{equation}
Squaring the leftmost and the rightmost sides and multiplying them by $D_{\alpha\beta}(E)$, we obtain
\begin{equation}
    \frac{\left(D_{\alpha \beta}(E) x^\nu_{\alpha}(E) (F_{\beta \alpha}^\pm(E) -F_{ \alpha\beta}^\pm(E))\right)^2}{D_{\alpha \beta}(E) (x^\nu_{\alpha}(E))^2 (F_{\beta \alpha}^\pm(E) +F_{ \alpha\beta}^\pm(E))} \leq D_{\alpha \beta}(E)  (F_{\beta \alpha}^\pm (E)+F_{ \alpha\beta}^\pm(E)). \label{eq: step 1 cl KUR}
\end{equation}
To find a bound in terms of average quantities, we apply Jensen's inequality, which states that for a function $\phi[x]$
\begin{equation}\label{eq:Jensen}
    \phi \left[ \frac{\sum_i \int ds w_i[s] y_i[s] }{\sum_i \int ds w_i[s] }\right] \lesseqgtr \frac{\sum_i \int ds w_i [s] \phi[y_i[s]]}{\sum_i \int ds w_i [s]}
\end{equation}
where $\leq$ applies if $\phi[y]$ is convex and $\geq$ applies if $\phi[y]$ is concave. We pick 
\begin{align}
    \phi[y] &= y^2,\\
    w_\beta[E] &=D_{\alpha \beta}(E) (x^\nu_{\alpha}(E))^2 (F_{\beta \alpha}^\pm(E) +F_{ \alpha\beta}^\pm(E)),\\
    y_\beta[E] &=D_{\alpha \beta}(E) x^\nu_{\alpha}(E) (F_{\beta \alpha}^\pm(E) -F_{ \alpha\beta}^\pm(E))/w_\beta(E),
\end{align}
for which Jensen's inequality implies
\begin{widetext}
\begin{align}
    \frac{\sum_{ \beta\neq\alpha} \int dE \left( \frac{D_{\alpha \beta}x_\alpha^\nu(F_{ \beta \alpha}^{\pm}-F_{\alpha \beta}^{\pm})}{D_{\alpha \beta}(x_\alpha^\nu)^2(F_{\alpha \beta}^{\pm}+ F_{ \beta \alpha}^{\pm})} \right)^2 D_{\alpha \beta} (x_\alpha^\nu)^2 (F_{\alpha \beta}^{\pm}+ F_{ \beta \alpha}^{\pm}) }{\sum_{ \beta\neq\alpha} \int dE  D_{\alpha \beta} (x_\alpha^\nu)^2 (F_{\alpha \beta}^{\pm}+ F_{ \beta \alpha}^{\pm})} &\geq  \left(  \frac{\sum_{ \beta\neq\alpha} \int dE D_{\alpha \beta}x_\alpha^\nu(F_{ \beta \alpha}^{\pm}-F_{\alpha \beta}^{\pm})}{\sum_{ \beta\neq\alpha} \int dE D_{\alpha \beta}(x_\alpha^\nu)^2(F_{\alpha \beta}^{\pm}+ F_{ \beta \alpha}^{\pm})} \right)^2 =\frac{\left( I^{(\nu)}_\alpha \right)^2}{\left(S_{\alpha \alpha}^{(\nu)\text{cl}} \right)^2 }. \label{eq: step 2 cl KUR}
\end{align}
\end{widetext}
Thus, by summing over reservoirs and integrating over energy on both sides of~\eqref{eq: step 1 cl KUR} we find that the precision with respect to the classical part of the fluctuations $\mathcal{P}_\alpha^{(\nu)\text{cl}}$ is limited by the classical particle current fluctuations 
\begin{equation}\label{app:eq: cl KUR}
     \mathcal{P}_\alpha^{(\nu)\text{cl}}\equiv S^{(N)\text{cl}}_{\alpha \alpha} \leq \frac{\left( I^{(\nu)}_\alpha \right)^2 }{S^{(\nu)\text{cl}}_{\alpha \alpha}} .
\end{equation}
Here, no assumptions about the scattering matrix or average occupations were made apart from the fact that the bosonic occupations fulfill $f_\alpha(E)\geq 0$ and that the fermionic ones fulfill the Pauli exclusion principle $1\geq f_\alpha(E) \geq0$. 

Following a similar calculation, it is also possible to derive a thermodynamic uncertainty relation for systems with time-reversal symmetry. Starting from the fact that the logarithmic mean is smaller than the arithmetic mean~\cite{Kwon2024Dec},
\begin{equation}\label{eq: step 1 TUR}
    \frac{F_{\beta \alpha}^{\pm}(E)-F_{\alpha \beta }^{\pm}(E)}{\log[F^\pm_{\beta\alpha }(E)]-\log[F^\pm_{\alpha \beta}(E)]} \leq \frac{F_{\beta \alpha}^{\pm}(E)+F_{\alpha \beta }^{\pm}(E)}{2}    
\end{equation}
it follows that
\begin{equation}\label{eq: step 2 TUR}
    \frac{\left( D_{\alpha \beta}(E) x^\nu_\alpha(E) \left(F_{\beta \alpha}^{\pm}(E)-F_{\alpha \beta }^{\pm}(E) \right)\right)^2}{D_{\alpha \beta}(E) (x^\nu_\alpha(E))^2 \left(F_{\beta \alpha}^{\pm}(E)+F_{\alpha \beta }^{\pm}(E)\right)} \leq \frac{1}{2}D_{\alpha \beta}(E) \log\left[\frac{F_{\beta \alpha}^{\pm}(E)}{F_{\alpha\beta }^{\pm}(E)} \right](F^\pm_{\beta\alpha}(E)-F^\pm_{\alpha\beta}(E)).
\end{equation}
{Furthermore, integrating over energy and summing over $\beta$ both sides of ~\eqref{eq: step 2 TUR} and applying~\eqref{eq: step 2 cl KUR} we find
\begin{equation}\label{eq: step 3 TUR}
    \frac{\left( I^{(\nu)}_{\alpha} \right)^2}{S^{(\nu)\mathrm{cl}}_{\alpha \alpha}} \leq \frac12 \sum_\beta \int \frac{dE}{h}D_{\alpha \beta} \log\left[\frac{F_{\beta \alpha}^{\pm}}{F_{\alpha\beta }^{\pm}} \right](F^\pm_{\beta\alpha}-F^\pm_{\alpha\beta}),
\end{equation}
where the right-hand side has terms that also appear in the symmetrized entropy production $\sigma^+$. In particular, using that $D_{\alpha\beta}(E)=D_{\beta\alpha}^\text{tr}(E)$, we have
\begin{equation}\label{eq: partial entropy production}
\begin{split}    
&\sum_\beta \int \frac{dE}{h}\left(\frac{D_{\alpha \beta}+D_{\beta\alpha}^\text{tr}}{2} \right) \log\left[\frac{F_{\beta \alpha}^{\pm}}{F_{\alpha\beta }^{\pm}} \right](F^\pm_{\beta\alpha}-F^\pm_{\alpha\beta}) =\\
&= \sum_\beta \int \frac{dE}{2h}D_{\alpha\beta} \log\left[\frac{F_{\beta \alpha}^{\pm}}{F_{\alpha\beta }^{\pm}} \right](F^\pm_{\beta\alpha}-F^\pm_{\alpha\beta})+ \sum_\beta \int \frac{dE}{2h}D_{\beta\alpha}^\text{tr} \log\left[\frac{F_{\alpha \beta}^{\pm}}{F_{\beta\alpha }^{\pm}} \right](F^\pm_{\alpha\beta}-F^\pm_{\beta\alpha}) \leq \frac{\sigma^+}{2\kB} + \frac{\sigma^+}{2\kB} = \frac{\sigma^+}{\kB},
\end{split}
\end{equation}
where the inequality stems from $\sigma^+$ containing also the (positive) contributions for each $\alpha$.
Then, the precision of the current flowing into reservoir $\alpha$ is limited by the symmetrized entropy production,
\begin{equation}
    \frac{\left( I^{(\nu)}_{\alpha} \right)^2}{S^{(\nu)\mathrm{cl}}_{\alpha \alpha}} \leq \frac12\frac{\sigma^+}{\kB}.
\end{equation}
Interestingly, taking the sum over $\alpha$ in Eq.~\eqref{eq: step 3 TUR} leads to
}
\begin{equation}
    \sum_\alpha \frac{\left( I^{(\nu)}_{\alpha} \right)^2}{S^{(\nu)\mathrm{cl}}_{\alpha \alpha}} \leq \frac{\sigma^+}{\kB},
\end{equation}
{where now the sum of the precisions of the currents flowing in \text{all} reservoirs is limited by the symmetrized entropy production.}
At last, we use~\eqref{eq: sigma plus bound} to extend the bound to systems with broken time-reversal symmetry
{
\begin{subequations}
\begin{align}
    \frac{\left( I^{(\nu)}_{\alpha} \right)^2}{S^{(\nu)\mathrm{cl}}_{\alpha \alpha}} &\leq \frac12\frac{\lambda \sigma}{\kB},\\
    \sum_\alpha \frac{\left( I^{(\nu)}_{\alpha} \right)^2}{S^{(\nu)\mathrm{cl}}_{\alpha \alpha}} &\leq \frac{\lambda \sigma}{\kB}.
\end{align}
\end{subequations}
}

\section{Deriviation of unified thermodynamic-kinetic uncertainty relation}\label{app: deriv TKUR}
In Ref.~\cite{Vo2022Sep}, the authors derived a unified thermodynamic-kinetic uncertainty relation (TKUR) for classical Markovian systems. Here we show the extension of this bound to quantum systems described by scattering theory. Mathematically, the derivations are similar, and we start from the equality~\cite{Vo2022Sep} 
\begin{equation}\label{eq: step 2 TKUR}
    \tanh\left[\frac{1}{2} \log [a/b] \right] = \frac{a-b}{b+a}\ .
\end{equation}
 Multiplying both sides of the equality by $\frac{1}{2} \log [a/b]$ and defining $\Omega(x)$ as the inverse of the function of $x \tanh{x}$ for $x\geq 0$, we have
\begin{equation}\label{eq: step 3 TKUR}
    1= \frac{1}{2} \log[a/b] \Omega^{-1} \left[ \frac{1}{2}\frac{(a-b)\log[a/b]}{a+b} \right].
\end{equation}
As a next step, we make a choice of $a$ and $b$ that is convenient for the scattering approach considered here. Concretely, we pick $a=F^\pm_{\beta \alpha}(E)$, $b=F^\pm_{\alpha\beta}(E)$ which allows us to write 
\begin{widetext}
    \begin{equation} \label{app:eq: step 4 TKUR}
         \frac{\left( D_{\alpha \beta} x_\alpha^{\nu} (F_{ \beta \alpha}^{\pm}-F_{\alpha \beta}^{\pm})\right)^2 }{D_{\alpha \beta} (x_\alpha^{\nu})^2 (F_{\alpha \beta}^{\pm}+ F_{ \beta \alpha}^{\pm})} = \frac{\left(\frac{1}{2} D_{\alpha \beta} (F_{ \beta \alpha}^{\pm}-F_{\alpha \beta}^{\pm}) \log\left[\frac{F_{\beta \alpha}^{\pm}}{F_{\alpha \beta}^{\pm}}\right] \right)^2 }{ D_{\alpha \beta }( F_{\alpha \beta}^{\pm} +F_{ \beta\alpha}^{\pm})} \Omega^{-2} \left[ \frac{\left( \frac{1}{2} D_{\alpha \beta}(F_{ \beta \alpha}^{\pm}-F_{\alpha \beta}^{\pm}) \log\left[\frac{F_{\beta \alpha}^{\pm}}{F_{\alpha \beta}^{\pm}}\right] \right) }{ D_{\alpha \beta }( F_{\alpha \beta}^{\pm}+F_{ \beta\alpha}^{\pm})} \right],
    \end{equation}
\end{widetext}
where we dropped the energy arguments for brevity. We are interested in taking a sum over reservoirs $\alpha,\beta$ and integrating over energy to relate Eq.~\eqref{app:eq: step 4 TKUR} to quantum transport quantities. The function on the right-hand side of Eq.~\eqref{app:eq: step 4 TKUR} $ x^2/(y \Omega^2[x/y])$ is concave for all $x,y\geq0$ \cite{Vo2022Sep} while the left-hand side is a convex function $x^2/y$. In order to use Jensen's inequality~\eqref{eq:Jensen}, we pick 
\begin{align}
    \phi[x] &= x^2 / \Omega^2[x],\\
    w_{\alpha \beta}[E] &=D_{\alpha \beta}(E)  (F_{\beta \alpha}^\pm(E) +F_{ \alpha\beta}^\pm(E)),\\
    y_{\alpha \beta}[E] &=\frac{D_{\alpha \beta}(E)(F_{ \beta \alpha}^{\pm}(E)-F_{\alpha \beta}^{\pm}(E))}{2w_{\alpha \beta}(E)} \log\left[\frac{F_{\beta \alpha}^{\pm}(E)}{F_{\alpha \beta}^{\pm}(E)}\right],
\end{align}
and perform sums over reservoirs and integrate over energy to find
\begin{widetext}
\begin{align}
     \frac{\sum_{\alpha \beta} \int dE  D_{\alpha \beta }( F_{\alpha \beta}^{\pm}+F_{ \beta\alpha}^{\pm}) \phi\left[ \frac{\left( \frac{1}{2}D_{\alpha \beta}(F_{ \beta \alpha}^{\pm}-F_{\alpha \beta}^{\pm}) \log\left[\frac{F_{\beta \alpha}^{\pm}}{F_{\alpha \beta}^{\pm}}\right] \right) }{ D_{\alpha \beta }( F_{\alpha \beta}^{\pm}+F_{ \beta\alpha}^{\pm})} \right]}{\sum_{\alpha \beta} \int dE  D_{\alpha \beta }( F_{\alpha \beta}^{\pm}+F_{ \beta\alpha}^{\pm})} &\leq \phi\left[ \frac{\frac{1}{2}\sum_{\alpha \beta} \int dE D_{\alpha \beta}(F_{ \beta \alpha}^{\pm}-F_{\alpha \beta}^{\pm}) \log\left[\frac{F_{\beta \alpha}^{\pm}}{F_{\alpha \beta}^{\pm}}\right] }{\sum_{\alpha \beta} \int dE D_{\alpha \beta }( F_{\alpha \beta}^{\pm}+F_{ \beta\alpha}^{\pm}) }\right] \nonumber \\
    &=\frac{1}{\kB^2} \frac{(\sigma^+)^2}{{\mathcal{K}^\text{cl}}^2} \Omega^{-2}\left[ \frac{1}{\kB} \frac{\sigma^+}{\mathcal{K}^\text{cl}} \right]. \label{eq: step 5 TKUR}
\end{align}
\end{widetext}
Here, we defined the total activity due to single-particle transfers as
\begin{equation}
    \mathcal{K}^\text{cl} \equiv \sum_\alpha S^{(N) \text{cl}}_{\alpha\alpha}.
\end{equation}
Applying~\eqref{eq: step 2 cl KUR} and~\eqref{eq: step 5 TKUR} to Eq.~\eqref{app:eq: step 4 TKUR} we arrive at the TKUR for the symmetrized entropy production
\begin{equation}\label{app:eq: TKUR cl}
    \sum_\alpha \frac{\left( I^{(\nu)}_\alpha \right)^2}{S_{\alpha \alpha}^{(\nu)\text{cl}}} \leq \frac{1}{\kB^2} \frac{(\sigma^+)^2}{\mathcal{K}^\text{cl}} \Omega^{-2}\left[ \frac{1}{\kB} \frac{\sigma^+}{\mathcal{K}^\text{cl}} \right].
\end{equation}
Furthermore, we use 
\begin{equation}
    \frac{d}{dx} \left(\frac{x^2 }{y \Omega^2(x/y)}\right) \geq 0
\end{equation}
for $x,y\geq 0$ allowing us to use~\eqref{eq: sigma plus bound} to extend the TKUR to systems with broken TRS
\begin{equation}\label{app:eq: TKUR cl}
    \sum_\alpha \frac{\left( I^{(\nu)}_\alpha \right)^2}{S_{\alpha \alpha}^{(\nu)\text{cl}}} \leq \frac{1}{\kB^2} \frac{(\lambda\sigma)^2}{\mathcal{K}^\text{cl}} \Omega^{-2}\left[ \frac{1}{\kB} \frac{\lambda\sigma}{\mathcal{K}^\text{cl}} \right].
\end{equation}

It is also possible to derive a version of the TKUR that only contains the local activity. To do so, we again start from Eq.~\eqref{app:eq: step 4 TKUR}, where we instead of taking the sum over both reservoirs, only take the sum over $\beta\neq\alpha$ and apply Jensen's inequality
\begin{align}\label{app:eq: TKUR local step 1}
    \frac{\left( I^{(\nu)}_\alpha \right)^2}{S_{\alpha \alpha}^{(\nu)\text{cl}}} \leq& \frac{\left(\int dE\sum_{\beta\neq\alpha} \frac{1}{2} D_{\alpha \beta}(F_{ \beta \alpha}^{\pm}-F_{\alpha \beta}^{\pm}) \log\left[\frac{F_{\beta \alpha}^{\pm}}{F_{\alpha \beta}^{\pm}}\right] \right)^2}{\mathcal{K}_\alpha^\text{cl}} \Omega^{-2}\left[ \frac{\left(\int dE \sum_{\beta\neq\alpha}\frac{1}{2} D_{\alpha \beta}(F_{ \beta \alpha}^{\pm}-F_{\alpha \beta}^{\pm}) \log\left[\frac{F_{\beta \alpha}^{\pm}}{F_{\alpha \beta}^{\pm}}\right] \right)}{\mathcal{K}_\alpha^\text{cl}} \right].
\end{align}
{
On the right-hand side, we have the same term appearing in~\eqref{eq: step 3 TUR}.
Then, using~\eqref{eq: partial entropy production} we extend the inequality to the symmetrized entropy production
}
\begin{equation}\label{app:eq: TKUR local step 2}
     \frac{\left( I^{(\nu)}_\alpha \right)^2}{S_{\alpha \alpha}^{(\nu)\text{cl}}} \leq \frac{1}{{4}\kB^2} \frac{(\sigma^+)^2}{\mathcal{K}_\alpha^\text{cl}} \Omega^{-2}\left[ \frac{1}{{2}\kB} \frac{\sigma^+}{\mathcal{K}_\alpha^\text{cl}} \right].
\end{equation}
Finally, we extend the bound to systems with broken TRS using~\eqref{eq: sigma plus bound}
\begin{equation}\label{app:eq: TKUR local step 2}
     \frac{\left( I^{(\nu)}_\alpha \right)^2}{S_{\alpha \alpha}^{(\nu)\text{cl}}} \leq \frac{1}{{4}\kB^2} \frac{(\lambda\sigma)^2}{\mathcal{K}_\alpha^\text{cl}} \Omega^{-2}\left[ \frac{1}{{2}\kB} \frac{\lambda\sigma}{\mathcal{K}_\alpha^\text{cl}} \right].
\end{equation}

Starting from~\eqref{app:eq: TKUR local step 1} we are also able to derive a similar bound to the TKUR with the intended goal of estimating $\sigma$. First, we note that we can replace the classical precision and local activity with their counterparts appearing in the local bounds for the full noise. Letting $\boldsymbol{\mathcal{P}}_\alpha$ and $\boldsymbol{\mathcal{K}}_\alpha$ denote any of the individual elements in the sums defining the short notations introduced in the Table~\ref{tab:boson_fermion_shorthand} and following the steps of the inversion of the TKURLs outlined in Sec.~\ref{sec:inference} we find 
\begin{widetext}
    \begin{equation}
        \frac{1}{2}\log\left[\frac{\sqrt{\boldsymbol{\mathcal{K}}_\alpha}+\sqrt{\boldsymbol{\mathcal{P}}_\alpha} }{\sqrt{\boldsymbol{\mathcal{K}}_\alpha}-\sqrt{\boldsymbol{\mathcal{P}}_\alpha}}\right]\sqrt{\boldsymbol{\mathcal{K}}_\alpha\boldsymbol{\mathcal{P}}_\alpha}\leq \int dE \sum_{\beta\neq\alpha} \frac{1}{2h} D_{\alpha \beta}(F_{ \beta \alpha}^{\pm}-F_{\alpha \beta}^{\pm}) \log\left[\frac{F_{\beta \alpha}^{\pm}}{F_{\alpha \beta}^{\pm}}\right].
\end{equation}
\end{widetext}
Taking the sum over both sides and using~\eqref{eq: sigma plus bound} we arrive at
\begin{equation}\label{app:eq:inference_bound2}
    \sum_\alpha\frac{1}{2}\log\left[\frac{\sqrt{\boldsymbol{\mathcal{K}}_\alpha}+\sqrt{\boldsymbol{\mathcal{P}}_\alpha} }{\sqrt{\boldsymbol{\mathcal{K}}_\alpha}-\sqrt{\boldsymbol{\mathcal{P}}_\alpha}}\right]\sqrt{\boldsymbol{\mathcal{K}}_\alpha\boldsymbol{\mathcal{P}}_\alpha}\leq\frac{\lambda\sigma }{\kB }.
\end{equation}
\subsection{Two-sided bounds on entropy production}\label{sec:two-sided}
While the TKUR and KUR are similar in the way they provide upper bounds on precision, they can also be combined to form an upper and lower bound on the total entropy production of a system, which we show here as a fun fact. Under the assumption that there are no hidden processes creating entropy, i.e., entropy production is captured by Eq.~\eqref{eq: entropy prod}, one can sum the local bosonic KURs~\eqref{eq: boson q TKUR} together with~\eqref{eq:inference_bound2} to get
\begin{equation}\label{eq:boson KURTUR}
  \sum_\alpha \sqrt{{S}^{(\Sigma)}_{\alpha\alpha} \frac{\widetilde{\mathcal{K}}_{\alpha,\text{bos}}}{1+\frac{\widetilde{\mathcal{K}}_{\alpha,\text{bos}}}{n_\alpha B^{\nu}_\alpha/h}}}\geq \sigma\geq \sum_\alpha\frac{\kB\sqrt{\boldsymbol{\mathcal{K}}_\alpha\boldsymbol{\mathcal{P}}_\alpha}}{2 \lambda}\log\left[\frac{\sqrt{\boldsymbol{\mathcal{K}}_\alpha}+\sqrt{\boldsymbol{\mathcal{P}}_\alpha} }{\sqrt{\boldsymbol{\mathcal{K}}_\alpha}-\sqrt{\boldsymbol{\mathcal{P}}_\alpha}}\right]
 \end{equation}
for bosonic systems. Analogously, using the fermionic KUR~\eqref{eq: fermion full q KUR} together with \eqref{eq:inference_bound2}, we find for fermionic systems
\begin{equation}\label{eq:fermion KURTUR}
    \sum_\alpha \frac{1}{R_\alpha}\sqrt{{S}^{(\Sigma)}_{\alpha\alpha} {S}^{(N)}_{\alpha\alpha}} \geq \sigma\geq \sum_\alpha\frac{\kB\sqrt{\boldsymbol{\mathcal{K}}_\alpha\boldsymbol{\mathcal{P}}_\alpha}}{2 \lambda}\log\left[\frac{\sqrt{\boldsymbol{\mathcal{K}}_\alpha}+\sqrt{\boldsymbol{\mathcal{P}}_\alpha} }{\sqrt{\boldsymbol{\mathcal{K}}_\alpha}-\sqrt{\boldsymbol{\mathcal{P}}_\alpha}}\right].
\end{equation}
This means that while activity and precision set a lower bound on the entropy production, particle- and entropy-current noise set an upper bound. 

\section{multi-channel bounds on quantum fluctuations.}

\subsection{Bosonic bound on quantum fluctuations} \label{app: bosonic q bounds}
We are interested in finding an upper bound on the quantum part of the bosonic fluctuations in order to make tight bounds for the full measurable fluctuations. To do this, we first define the hermitian matrix
\begin{equation}
    \mathcal{G}_\alpha(E) = \sum_{\beta\neq\alpha} t_{\alpha \beta}(E) t_{\alpha \beta}^\dag(E)  (f_\beta(E)-f_\alpha(E))
\end{equation}
and use the Cauchy-Schwarz inequality for the Hilbert-Schmidt product of matrices to establish
\begin{align} \label{eq: step 1 Boson Q bound}
    n_\alpha S^{(\nu)\text{qu}}_{\alpha \alpha,\text{bos}}=& \frac{1}{h} \int dE \text{tr}\left\{\mathcal{G}_\alpha^2\right\}\text{tr}\left\{\mathbb{I}_{\alpha}^2\right\}(x^\nu_\alpha)^2 \geq \frac{1}{h}  \int dE\text{tr}\left\{\mathcal{G}_\alpha\right\}^2(x^\nu_\alpha)^2.
\end{align}
Here, $n_\alpha$ is the number of channels in the lead connected to reservoir $\alpha$. Next, we define the indicator function of the spectral current 
\begin{equation}
\zeta_\alpha^\nu (E)=
    \begin{cases}
         1 \quad \text{if} \quad E\in\text{supp}\{x_\alpha^{(\nu)}(E)\sum_\beta D_{\alpha\beta}(E)\left(f_\beta(E)-f_\alpha(E)\right)\},  \\
         0 \quad \text{otherwise}
    \end{cases}
\end{equation}
together with the bandwidth of transport 
\begin{equation}
    B_\alpha^\nu = \int dE \zeta_\alpha^\nu (E).
\end{equation}
Using the Cauchy-Schwarz inequality for square integrable functions, we find
\begin{equation} \label{eq: step 2 Boson Q bound}
    \frac{1}{h}\int dE \text{tr}\left\{\mathcal{G}_\alpha\right\}^2(x^\nu_\alpha)^2 \geq \frac{h}{B_\alpha^ \nu}\left|I^{(\nu)}_{\alpha}\right|^2.
\end{equation}
Combining~\eqref{eq: step 1 Boson Q bound} with~\eqref{eq: step 2 Boson Q bound} we bound the quantum part of the fluctuations 
\begin{equation}\label{app: eq: Boson Q bound}
    \frac{n_\alpha B^{\nu}_\alpha}{ h} \geq \frac{\left({I^{(\nu)}_{\alpha}}\right)^2 }{ S^{(\nu)\text{qu}}_{\alpha \alpha}} \geq \frac{\left({I^{(\nu)}_{\alpha}}\right)^2 }{ S^{(\nu)}_{\alpha \alpha}}. 
\end{equation}

\subsection{Fermionic bound on quantum fluctuations} \label{app: fermionic q bounds}
Dropping energy arguments and using the cyclicity of the trace, we note that the trace of the four scattering submatrices is positive
\begin{equation}
\text{tr}\left\{t_{\alpha \beta} t_{\alpha \beta}^\dag t_{\alpha \gamma} t_{\alpha \gamma}^\dag\right\}=\text{tr}\left\{t_{\alpha \gamma}^\dag t_{\alpha \beta} \left(t_{\alpha \gamma}^\dag t_{\alpha \beta}\right)^\dag  \right\} \geq 0.
\end{equation}
Moreover, using the property of fermionic average occupations $(f_\beta-f_\alpha)\in[-1,1]$ and the unitarity of the scattering matrix, we find
\begin{equation}
\begin{split}
   \sum_{\beta,\gamma\neq\alpha} \text{tr}\left\{t_{\alpha \beta} t_{\alpha \beta}^\dag t_{\alpha \gamma} t_{\alpha \gamma}^\dag\right\} (x^\nu_\alpha)^2 (f_\beta-f_\alpha)(f_\gamma-f_\alpha) &\leq    \sum_{\beta,\gamma\neq\alpha} \text{tr}\left\{t_{\alpha \beta} t_{\alpha \beta}^\dag t_{\alpha \gamma} t_{\alpha \gamma}^\dag\right\} (x^\nu_\alpha)^2 \left|(f_\beta-f_\alpha)(f_\gamma-f_\alpha)\right|\\
   &\leq \sum_{\beta,\gamma\neq\alpha} \text{tr}\left\{t_{\alpha \beta} t_{\alpha \beta}^\dag t_{\alpha \gamma} t_{\alpha \gamma}^\dag\right\} (x^\nu_\alpha)^2 |f_\gamma-f_\alpha| \\
   &= \sum_{\gamma\neq\alpha} \text{tr}\left\{\left(\mathbbm{1}_\alpha - t_{\alpha\alpha}t_{\alpha\alpha}^\dagger\right) t_{\alpha \gamma} t_{\alpha \gamma}^\dag\right\} (x^\nu_\alpha)^2 |f_\gamma-f_\alpha|.
\end{split}
\end{equation}
The reflection probabilities of separate channels $i$ in $\alpha$ are given by the eigenvalues $t_{\alpha\alpha}(E)t_{\alpha\alpha}^\dagger(E)\ket{\alpha i}=R_{\alpha,i}(E)\ket{\alpha i}$. We can use this to write
\begin{equation}\label{eq: fermionic QN bound}
\begin{split}
   &\sum_{\beta,\gamma\neq\alpha} \text{tr}\left\{t_{\alpha \beta} t_{\alpha \beta}^\dag t_{\alpha \gamma} t_{\alpha \gamma}^\dag\right\} (x^\nu_\alpha)^2 (f_\beta-f_\alpha)(f_\gamma-f_\alpha) \leq \left(1-\min_i R_{\alpha,i} \right)  \sum_{\gamma\neq\alpha} \text{tr}\left\{t_{\alpha \gamma} t_{\alpha \gamma}^\dag\right\} (x^\nu_\alpha)^2 |f_\gamma-f_\alpha|.
\end{split}
\end{equation}
Defining the minimum reflection probability $R_\alpha = \min_i \inf_{E \in A}R_{\alpha,i}(E) $ where $A$ is the support of the spectral current and using
\begin{equation}
    D_{\alpha\beta}\left|f_\beta-f_\alpha\right|=D_{\alpha\beta}\left|F_{\beta \alpha}^\pm -F_{ \alpha\beta}^\pm \right|\leq D_{\alpha\beta}\left(F_{\beta \alpha}^\pm +F_{ \alpha\beta}^\pm\right),
\end{equation}
see also~\eqref{eq:f_inequality} in the main text, we find
\begin{equation}
    (1-R_\alpha) S^{(\nu)\text{cl}}_{\alpha\alpha} \geq - S^{(\nu)\text{qu}}_{\alpha\alpha}.
\end{equation}
This implies that
\begin{equation}\label{app: eq: fermionic q bound}
 \frac{S^{(\nu)}_{\alpha\alpha}}{R_\alpha} \geq S^{(\nu)\text{cl}}_{\alpha\alpha}.
\end{equation}

\section{Mathematical properties of $\Omega[x]$}\label{app: omega}
\begin{figure}[h!]
    \centering
    \includegraphics[width=0.5\columnwidth]{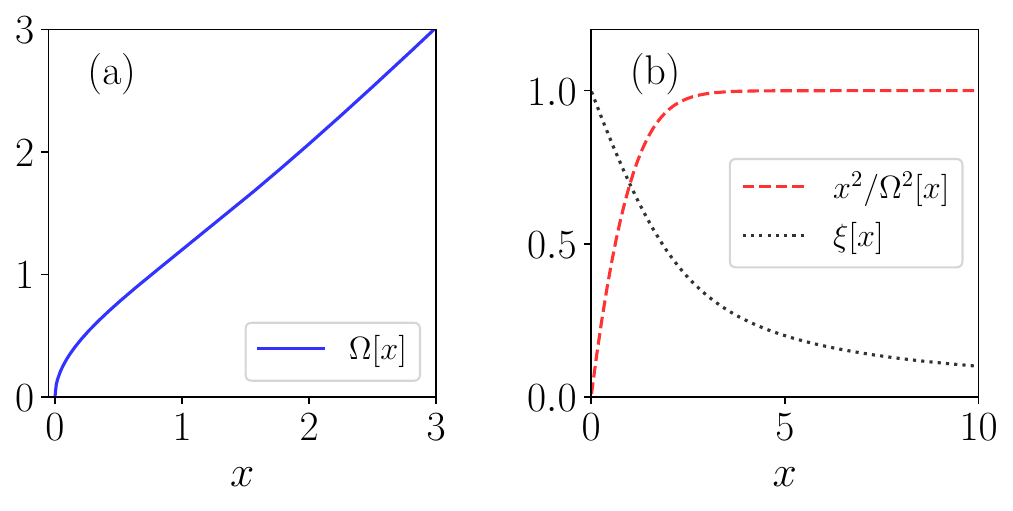}
    \caption{Behaviour of the functions $\Omega[x]$ defined as the inverse of $x\tanh x$ and $\xi[x] =x/\Omega^2[x]$ present in our bounds.}  
    \label{fig:functions}
\end{figure}
In this section, we prove two useful properties of the function $x^2 /(y \Omega^2[x/y])$, see Eq.~\eqref{eq:xi} and the paragraph around it, that we apply in order to write TKURLs for systems with broken TRS and in terms of full fluctuations. We start by proving
\begin{equation}\label{app eq: Omega dx}
    \frac{d}{dx} \frac{x^2}{y \Omega^2[x/y]} \geq0
\end{equation}
for $x,y\geq 0$. By using the substitution, $u[x] =\Omega[x/y]$ we are able to write,
\begin{equation}
    \frac{x^2}{y \Omega^2[x/y]} = y \tanh^2[u[x]]
\end{equation}
which in turn implies that
\begin{align}
     &\frac{d}{dx} \frac{x^2}{y \Omega^2[x/y]} =  y \left(\frac{d}{du} \tanh^2[u[x]] \right) \left(\frac{d}{dx}u[x]\right) y \,\mathrm{sech}^2[u[x]]\tanh[u[x]] \left(\frac{d}{dx}\Omega[x/y] \right)\geq0,
\end{align}
since $x \tanh[x]$ is a strictly increasing function of $x$. This means that its inverse $\Omega[x]$ is also strictly increasing, proving~\eqref{app eq: Omega dx}. Next we prove
\begin{equation}\label{app eq: Omega dy}
    \frac{d}{dy} \frac{x^2}{y \Omega^2[x/y]} \geq0.
\end{equation}
We again use the substitution $u[y] = \Omega[x/y]$ meaning that
\begin{align}
     &\frac{d}{dy} \frac{x^2}{y \Omega^2[x/y]} = \frac{d}{dy}\left(x \frac{\tanh[u[y]]}{u[y]} \right)= x \left( \frac{d}{du} \frac{\tanh[u[y]]}{u[y]} \right) \left( \frac{d}{dy} \Omega[x/y]\right). \nonumber
\end{align}
Furthermore, we note that
\begin{equation}
    \frac{d}{du} \frac{\tanh[u]}{u} =\frac{u \,\mathrm{sech}^2[u]-\tanh[u]}{u^2}\leq0
\end{equation}
since
\begin{equation}
    u \,\mathrm{sech}^2[u]\leq\tanh[u] \Longleftrightarrow 2u \leq \sinh[2u].
\end{equation}
Finally we again use $\frac{d}{dx} \Omega[x]\geq0$ meaning that $\frac{d}{dy} u[y] =\frac{d}{dy}\Omega[x/y]\leq0$ proving~\eqref{app eq: Omega dy} for $x,y\geq0$.

\bibliography{refs.bib}
\end{document}